\documentclass[a4paper,10pt]{article}
\usepackage{graphicx} 
\usepackage[utf8]{inputenc}
\usepackage[T1]{fontenc}
\usepackage{comment}
\usepackage[T1]{fontenc}
\usepackage{subfigure}
\usepackage{float}
\usepackage{natbib}
\usepackage{epstopdf}
\usepackage{xcolor}
\usepackage{a4wide}
\usepackage{indentfirst}
\usepackage{authblk}

\usepackage{tikz}
\usetikzlibrary{patterns}
\usetikzlibrary{arrows.meta,positioning}
\usepackage{standalone}

\usepackage{amsmath,amssymb}
\usepackage{amsthm}
\usepackage{hyperref}
\usepackage{mathtools}

\usepackage[margin=1.2in]{geometry}
\usepackage{graphicx}
\usepackage{subcaption}
\usepackage{listings}
\usepackage[utf8]{inputenc}
\usepackage{appendix}
\numberwithin{equation}{section}
\newtheorem{teo}{Theorem}[section]
\newtheorem*{teo*}{Theorem}
\newtheorem*{prop*}{Proposition}
\newtheorem*{corol*}{Corollary}
\newtheorem{prop}[teo]{Proposition}

\newtheorem{lema}[teo]{Lemma}
\newtheorem{defi}[teo]{Definition}
\newtheorem{hyp}{Hypothesis}

\theoremstyle{definition}

\newtheorem{remark}[teo]{Remark}

\newcommand{\BS}{\operatorname{BS}}
\newcommand{\B}{\mathcal{B}}

\renewcommand{\d}{\,\mathrm{d}}

\newcommand{\D}{\mathbb{D}}
\newcommand{\E}{\mathbb{E}}
\newcommand{\F}{\mathcal{F}}

\newcommand{\N}{\mathbb{N}}
\renewcommand{\P}{\mathbb{P}}

\newcommand{\R}{\mathbb{R}}

\newcommand{\1}{\mathbf{1}}
\newcommand{\phy}{\varphi}

\title{On the skew-curvature power law relationship in a stochastic volatility model with infinite variation Lévy jumps}
\author[1]{Òscar Burés}
\author[1,2]{Josep Vives}

\affil[1]{ Universitat de Barcelona, Departament de Matemàtica Econòmica, Financera i Actuarial. \authorcr Diagonal 690--696, 08034 Barcelona, Spain, \vspace*{3pt}}
\affil[2]{Institut de Matemàtiques de la Universitat de Barcelona. \authorcr Gran Via de les Corts Catalanes, 585, 08007 Barcelona, Spain}

\date{\today}
\date{\today}

\begin{document}

\maketitle

\begin{abstract}
The short-term blow up behaviour of the ATM-IV skew is commonly modelled by either a rough volatility process with small Hurst index or an infinite variation Lévy process in the log-price dynamics. In this paper, we examine which is the effect of mixed dynamics (i.e. rough volatility and infinite variation Lévy processes) on the ATM-IV skew and curvature.

The results of the paper show that the analysis of the leading terms of the short-time expansion of both ATM-IV skew and curvature allow us to determine, from a mixed dynamics model, which mechanism originates the power-law behaviour on the skew and on the curvature. Apart from showing that the mixed model also originates a power law, it has also key implications in model design and calibration.

Numerical evidences for the rough Heston model with CGMY Lévy jumps are provided.

\end{abstract}


\tableofcontents

\section{Introduction}
In quantitative finance, option pricing is one of the central problems, and the implied volatility surface is one of the most interesting areas of study within option pricing. One of the reasons why the study of implied volatility is so important is that, in certain regions of the surface, we are able to find information about the risk-neutral distribution of the asset and the volatility process. Therefore, the implied volatility surface plays a fundamental role in model construction, identification, and calibration. In particular, the presence of smiles shows that the classical Black–Scholes model (see \cite{BlackScholes1973}), which assumes constant volatility, cannot reproduce observed option prices since the resulting implied volatility surface is a plane. Local volatility models, introduced in \cite{Dupire1994}, can fit a given implied volatility surface but they may not capture satisfactorily the smile dynamics.

A classical answer to overcome such limitations is to model volatility as a stochastic process, originating the so-called stochastic volatility models (see, for instance \cite{HullWhite1987}, \cite{Scott1987}, \cite{SteinStein1991}, \cite{Heston1993} and \cite{HaganEtAl2002}, among many others.

Although stochastic volatility models generate non-trivial implied volatility smiles, they produce continuous stock-price trajectories and therefore are unable to reproduce discontinuities on the price dynamics. Jump-diffusion models address this issue by introducing a compound Poisson process into the log-price dynamics. The \cite{Bates1996} model, which combines the CIR stochastic variance process with a compound Poisson process, is the fundamental example of models of such kind. The resulting model simultaneously incorporates stochastic volatility, leverage, and discontinuous stock-price movements.

These models considerably improve the flexibility of the implied volatility surface and the realism of the underlying price dynamics. Nevertheless, finite activity Lévy process and conventional Markovian stochastic volatility  models do not reproduce an important feature of short-dated options: the power-law behaviour of the at-the-money implied volatility skew. Let $I(T,k)$ denote the implied volatility as a function of log-strike $k$ and maturity $T$. Empirical studies indicate that the ATM-IV skew $\partial_k I(T,k^*)$ (here $k^*$ denotes the at-the-money log-strike) becomes increasingly steep as maturity approaches zero, approximately following a power law of order $T^{-0.4}$. See \cite{Fukasawa2017}, \cite{GatheralJaissonRosenbaum2018}. This observation has become an important criterion for model construction.

Two main mechanisms have been classically proposed to reproduce this power law. The first is rough volatility models (see, for instance \cite{GatheralJaissonRosenbaum2018}). In rough volatility models, volatility is modelled by a function of a fractional or Volterra-type process with Hurst parameter $H\in(0,1/2)$. These models usually generate a power law on the ATM skew of order $T^{H-1/2}$, which agrees with the empirical rate when $H$ is close to $0.1$, as discussed in \cite{Fukasawa2017} and  \cite{ElEuchFukasawaGatheralRosenbaum2019}. Even though rough volatility models are non-Markovian, they produce continuous dynamics.

The second mechanism is provided by infinite variation Lévy processes, which exhibit infinitely many small jumps. Classical infinite variation Lévy processes used in financial modelling are, for instance, the Variance Gamma model of \cite{MadanCarrChang1998} and the CGMY model introduced by \cite{CarrGemanMadanYor2002}. For stable-like Lévy processes (introduced in \cite{FL2016} and \cite{FL2016-2}) with activity index $Y\in(1,2)$, including CGMY-type Lévy measures, the infinite-variation jump component can generate an ATM skew of order $T^{(1-Y)/2}$. Hence, values of $Y$ close to $1.8$ also reproduce the empirically observed rate. Rough volatility and infinite-variation jumps thus provide two explanations for the same power-law behaviour detected in the ATM-IV skew. 

The two mechanisms have been studied extensively in isolation. Short-time expansions for the ATM skew under continuous rough volatility models were obtained, among others, by \cite{Fukasawa2017} and \cite{ElEuchFukasawaGatheralRosenbaum2019}. For rough volatility models with a compound Poisson process, the authors in \cite{ALV} provide a key result that shows the joint contribution of the stochastic volatility and the jumps to the ATM-IV skew. In a different direction, \cite{FigueroaLopezOlafsson2016} derived high-order expansions for the ATM skew under a broad class of stochastic volatility models with independent stable-like jumps of infinite variation. Results combining rough volatility and infinite-variation jumps are scarce. A particularly relevant contribution is the one in \cite{FordeSmithViitasaari2021}, where the authors derive an Edgeworth-type expansion for near-the-money option prices under a rough the rough Heston model and a CGMY jump component.

Compared to the literature on the ATM-IV skew, the amount of results in the short-term behaviour of the ATM-IV curvature is relatively scarce. In \cite{AlosLeon2017}, the authors obtain a representation for the curvature in a general stochastic volatility model using Malliavin calculus. Similarly, in \cite{AlosLorite2023}, the authors compare the ATM-IV curvature for local and implied volatilities. In \cite{ElEuchFukasawa2019} the authors derive joint skew and curvature asymptotics by means of an Edgeworth expansion. ATM-IV curvature asymptotics for infinite-variation Lévy processes are much less developed compared to the literature on ATM-IV skew under infinite variation Lévy process and ATM-IV curvature results under stochastic volatility models.     

In this paper, we provide a general characterization of the leading term in the short-time expansion of the ATM implied volatility skew and curvature under a general stochastic volatility model and stable-like Lévy processes. The volatility framework includes both Markovian stochastic volatility models cited above and fractional or Volterra-type rough volatility.

Our results reveal two distinct phase transitions. The skew is dominated by rough volatility when $H<1-Y/2$ and by the Lévy process when $H>1-Y/2$. By contrast, the curvature changes its dominant mechanism at the threshold $Y = 2-4H$. The two thresholds divide the parameter space into three possible regimes: both the skew and curvature are dominated by rough volatility, the skew is dominated by rough volatility while the curvature is dominated by jumps, or both quantities are dominated by jumps. The possible regime in which jumps dominate the skew while rough volatility dominates the curvature, is impossible to happen. This phase structure has direct consequences for model identification. When rough volatility or infinite-variation jumps are considered in isolation, the skew and curvature powers are constrained by an algebraic relationship. In the combined model, these rigid relations can be broken, leading to more model flexibility for fitting the ATM-IV skew and curvature.

The proofs of the results rely on Malliavin calculus techniques. The general theorems are initially established under standard Malliavin regularity and non-degeneracy assumptions. Although these assumptions are convenient for the analysis, they are not directly satisfied by several important rough volatility models. We show how this limitation can be overcome through an approximation argument. In particular, for the rough Heston model (see \cite{ElEuchRosenbaum2019}), we construct an approximating volatility process satisfying the hypotheses of the general theorems and prove that it has the same leading short-time asymptotics as the original model. This transfers the conclusions rigorously to rough Heston and provides a methodology that may be applicable to a deeper study of the implied volatility under general models.

The paper is organized as follows. Section \ref{sec 2} introduces the general model and states the main results (i.e. Theorem \ref{th: main result} and Theorem \ref{th: main result 2} under conditions on the stochastic volatility process and the Lévy measure. Section \ref{sec 3} contains the proof of Theorem \ref{th: main result}. Section \ref{sec 4} contains the proof of Theorem \ref{th: main result 2}. Finally, section \ref{sec 5} develops the approximation argument for the rough Heston model and presents numerical evidence supporting the theoretical results.

\section{Statement of the model} \label{sec 2}
Throughout the paper we will assume, for simplicity, zero interest rate $r = 0$. Let $T > 0$ be a maturity time. We will consider two different models for an asset $S$. On the one hand, we will assume that $S$ follows a classical log-normal model with stochastic volatility with jumps, whose log-price dynamics can be written as
\begin{equation} \label{Black model}
X_t = X_0 - \eta t - \frac{1}{2}\int_0^t \sigma_s^2 ds + \int_0^t \sigma_s (\rho dW_s + \sqrt{1-\rho^2}dB_t) + L_t, \quad S_t = e^{X_t}.
\end{equation}
 Here, $W$ and $B$ denote independent Brownian motions, $\rho \in (-1,1)$, $\sigma$ is a stochastic process with continuous trajectories adapted to the filtration generated by $W$ and $L$ denotes a Lévy process with Lévy measure $\nu$ of the form
 \[
 L_t = \int_0^t \int_{|y| > 1} y N(ds,dy) + \int_0^t \int_{|y| \leq 1} y \tilde{N}(ds,dy), \quad \tilde{N}(ds,dy) = N(ds,dy) - ds\nu(dy).
 \]
 In order for $e^{X_t}$ to be a martingale, we need
 \[
 \eta = \int_{\R} (e^y - 1- y\1_{\{|y| \leq 1\}})\nu(dy), \quad \int_{y > 1} e^{y}\nu(dy) < \infty.
 \]
 All the stochastic processes involved are defined in the same filtered probability space $(\Omega, \mathcal{F}, \mathbb{F}, \P)$, where $\mathbb{F}$ denotes the filtration generated by $W$, $B$ and $L$.

 If the volatility process is constant and there are no jumps (i.e. $\sigma \equiv \sigma_0$ and $L \equiv 0$), then model \eqref{Black model} becomes the Black-Scholes model.
 
 Let $\BS$ denote the Black-Scholes function, that is,
 \[
 \BS(T,t,x,k,\sigma) = e^{x}\Phi(d_+) - e^{k} \Phi(d_{-}), \quad d_{\pm} = \frac{x-k}{\sigma \sqrt{T-t}} \pm \frac{\sigma}{2}\sqrt{T-t}.
 \]
 The price of a Call option with maturity $T$ and strike $K = e^{k}$ at time $t = 0$ under the Black-Scholes model can be computed as
 \[
 \mathcal{C}(T,k)  = \E[(e^{X_T} - e^k)_+] = \BS(T,0,X_0,k, \sigma).
 \]
 When the option is at-the-money, that is, $k = X_0$, we will denote the atm-strike as $k^* = X_0$. We thus simplify the previous notation to
 \[
 \mathcal{C} := \mathcal{C}(T,k^*) = \BS(T,0,X_0, \sigma).
 \]
Since $\partial_{\sigma}\BS(T,T,t,x,k,\sigma) > 0$, we can define the implied volatility as the unique quantity $I \equiv I(T,k)$ such that, for an observed market Call price $\mathcal{C}(T,k)$, the relationship
\[
\mathcal{C}(T,k) = \BS(T,0,X_0, k, I(T,k))
\]
is satisfied. We write the at-the-money implied volatility $I(T,k^*)$ as $I(T,k^*) = \BS^{-1}(T,0,X_0, \mathcal{C})$, where $\BS^{-1}$ denotes the inverse of the $\BS$ function with respect to $\sigma$ and $\mathcal{C} = \mathcal{C}(T,k^*)$ denotes the ATM call price. If we assume $I(T,k)$ to be differentiable at least once with respect to the log-strike $k$, we can then do a Taylor expansion around the ATM log-strike $k^* = X_0$ to obtain the first-order approximation
\[
I(T,k) \approx I(T,k^*) + \partial_k I(T,k^*)(k-k^*).
\]
In this paper, we are interested in the study of the quantity $\partial_k I(T,k^*)$ as $T \to 0$. This paper generalises significantly the results in \cite{ALV}. In such paper, the authors work with $\partial_{X_0} I(T,k^*)$ instead of $\partial_k I(T,k^*)$. The relationship between those quantities can be easily established by the well-known fact that
\[
\partial_{X_0} I(T,k^*) = -\partial_k I(T,k^*).
\]
\subsection{Malliavin calculus tools}
This section is devoted to provide a primer on the Malliavin calculus tools we have used in the paper. For a complete reference, we refer the reader to \cite{nualart2006malliavin}. Let $W = \{W_t; t \in [0,T]\}$ be a standard Brownian motion defined in a complete probability space $(\Omega, \F, \mathbb{P})$. For a function $h \in L^2([0,T])$ we define $W(h)$ as the Wiener integral of $h$ with respect to $W$, i.e.
\[
W(h) := \int_0^T h(t) \d W_t.
\]
We define $\mathcal{S}$ as the class of random variables of $F$ the form
\[
F = f(W(h_1), \dots, W(h_n)); \quad h_i \in L^2([0,T]), \quad f\in \mathcal{C}^{\infty}_b(\R^n),
\]
where $\mathcal{C}_b^{\infty}(\R^n)$ denotes the space of continuous and bounded functions which are infinitely many times differentiable and $f$, together with its partial derivatives, are continuous and bounded. We say that $\mathcal{S}$ is the space of \textit{simple functionals}. 
\begin{defi}
    Let $F \in \mathcal{S}$. We define the Malliavin derivative of $F$ evaluated at a point $t \in [0,T]$ as the stochastic process defined by
    \[
    D_t F = \sum_{j=1}^n \partial_j f(W(h_1), \dots, W(h_n)) h_j(t), \quad \partial_j f := \frac{\partial f}{\partial x_j}.
    \]
\end{defi}
As it has been proved in \cite{nualart2006malliavin}, the operator $D$ is closable from $L^p(\Omega)$ to $L^p(\Omega \times [0,T])$ for all $p \geq 1$. We define $\D^{1,p}_{W}$ as the closure of $\mathcal{S}$ with respect to the norm
\[
||F||_{1,p} = \left[ \E(|F|^p) + \E\left(||DF||_{L^2([0,T])}^p\right)\right]^{1/p}.
\]
In the same way we have defined $D$, we can define the iterated derivatives $D^n$. We denote $\D^{n,p}_{W}$ the space of random variables that are $n$ times Malliavin differentiable and $F$, together with its Malliavin derivatives are $p$-integrable.

\subsection{Hypotheses and main results}
In this section we will assume a set of hypotheses for the Lévy measure $\nu$ and the volatility process $\sigma$.
\begin{hyp} \label{hyp 1}
    The Lévy measure $\nu$ is of the form
    \[
    \nu(dy) = C\left(\frac{y}{|y|} \right)q(y) |y|^{-1-Y}dy, \quad Y \in (1,2)
    \]
    where $q$ is a smooth function on $(0,\infty)$ and on $(-\infty,0)$ such that $q(y) \to 1$ as $y \to 0$. Assume also that
    \[
    \int_{|y| > 1} |y| \nu(dy) <  \infty.
    \]We also write $C_+ = C(1)$ and $C_{-} = C(-1)$.
\end{hyp}
\begin{hyp} \label{hyp 2}
    There exist $0 < \alpha < \beta$ such that $\alpha \leq \sigma_t \leq \beta$ for almost every $t \in [0,T]$.
\end{hyp}
\begin{hyp} \label{hyp 3}
    For all $p \geq 2$, $\sigma_t \in \D^{2,p}_W$ for almost every $t \in [0,T]$.
\end{hyp}
\begin{hyp} \label{hyp 4}
    There exists $H \in (0,1)$ and, for all $p \geq 1$, there exist constants $\gamma,  \delta> 0$ such that for almost every $0 \leq s \leq r\leq u \leq T$ and for almost every $t$,
    \[
    \E\left( |D_r\sigma_u^2|^p\right)^{1/p} \leq \gamma(u-r)^{H-1/2},
    \]
    and
    \[
    \E\left( |D_s D_r \sigma_u^2|^p\right)^{1/p} \leq \delta (u-r)^{H-1/2}(u-s)^{H-1/2}.
    \]
\end{hyp}
\begin{hyp} \label{hyp 5}
    The process $\sigma$ has continuous trajectories and, moreover,
    \[
    \lim_{T \to 0} \sup_{r,s \in [0,T]} \E[|\sigma_s\sigma_r - \sigma_0^2|] = 0
    \]
\end{hyp}
\begin{hyp} \label{hyp 6}
    There exists a constant $\Sigma \geq 0$ such that
    \[
    \lim_{T \to 0} \E\left[ \left|\frac{D_s\sigma_r^2}{(r-s)^{H-1/2}} - \Sigma\right|\right] = 0.
    \]
    That is, $D_s\sigma_r^2 = \Sigma(r-s)^{H-1/2} + o((r-s)^{H-1/2})$ in $L^1(\Omega)$.
\end{hyp}
Under these hypotheses we can state the first result of our paper, which characterizes the leading power-law behaviour of the ATM-IV skew.
\begin{teo} \label{th: main result}
    Let $X$ follow the model \eqref{Black model}. Assume that the Lévy process $L$ and the volatility process $\sigma$ satisfy Hypotheses \ref{hyp 1} -- \ref{hyp 6}. Then, the ATM implied volatility skew, $\partial_k I(T,k^*)$, exhibits the following short-time behaviour:
\[\lim_{T\to0}
\frac{\partial_k I(T,k^*)}
{T^{\min\left(H-\frac12,\frac{1-Y}{2}\right)}}
=
\begin{cases}
\dfrac{\rho\Sigma}
{2\sigma_0^2\left(H+\frac12\right)\left(H+\frac32\right)},
& \text{if } H-\frac12<\frac{1-Y}{2}, \\[1.2em]

\dfrac{(C_+-C_-)\Gamma\left(\frac{3-Y}{2}\right)}
{\sigma_0^Y Y(Y-1)2^{\frac{Y-1}{2}}},
& \text{if } H-\frac12>\frac{1-Y}{2}, \\[1.2em]

\dfrac{\rho\Sigma}
{2\sigma_0^2\left(H+\frac12\right)\left(H+\frac32\right)}
+
\dfrac{(C_+-C_-)\Gamma\left(\frac{3-Y}{2}\right)}
{\sigma_0^Y Y(Y-1)2^{\frac{Y-1}{2}}},
& \text{if } H-\frac12=\frac{1-Y}{2}.
\end{cases}
\]
\end{teo}
In Section \ref{sec 3} we will give the proof of Theorem \ref{th: main result}.
\begin{remark}
    In this paper we restrict ourselves to the case $H \in (0,1/2)$ and $Y \in (1,2)$ because the power-law behaviour for general values of $H$ can be easily deduced. Indeed, as a consequence of our result, if $H > \frac{1}{2}$ the ATM-IV skew satisfies the following short-time behaviour
        \[
        \partial_k I(T,k^*) \sim - \frac{(C_{-}-C_+)\Gamma\left(\frac{3-Y}{2}\right)}{\sigma_0^Y Y(Y-1)2^{\frac{Y-1}{2}}} T^{\frac{1-Y}{2}}.
        \]
\end{remark}
\begin{remark}
We can also extend the result for $Y \in (0,1)$. In this case, we can adapt the results in \cite{ABV2026} to the log-normal context to ensure that
        \[
        \partial_kI(T,k^*) \sim \begin{cases}
        \displaystyle
            \dfrac{\int_{\R} (e^y - 1) \nu(dy)}{\sigma_0} + \lim_{T \to 0} \dfrac{\rho}{\sigma_0 T^2}\int_0^T \int_t^T\E[D_t\sigma_s] ds dt & H \geq 1/2, \\
            \displaystyle
            \dfrac{\rho}{\sigma_0 T^{3/2 + H}}\int_0^T \int_t^T\E[D_t\sigma_s] ds dt & H < 1/2.
        \end{cases}
        \]
\end{remark}
By strengthening our hypotheses on the Malliavin derivatives of the volatility process $\sigma$ we are able to obtain an analogous result for the ATM-IV curvature.
\begin{hyp} \label{hyp 7}
    The volatility process satisfies $\sigma_t \in \D^{3,p}$ for almost every $t \in [0,T]$.
\end{hyp}
\begin{hyp} \label{hyp 8}
    There exists a constant $\Theta$ such that for all $0 \leq r \leq s \leq t \leq T$,
    \[
    \lim_{T \to 0} \E\left[\left| \frac{D_rD_s\sigma^2_t}{(t-s)^{H-\frac{1}{2}}(s-r)^{H-\frac{1}{2}}}- \Theta\right|\right] = 0.
    \]
    That is, $D_rD_s\sigma_t^2 = \Theta(t-s)^{H-\frac{1}{2}}(s-r)^{H-\frac{1}{2}} + o((t-s)^{H-\frac{1}{2}}(s-r)^{H-\frac{1}{2}})$ in $L^1(\Omega)$.
\end{hyp}
\begin{hyp} \label{hyp 9}
    There exists $H \in (0,1)$ and, for all $p \geq 1$, there exists a constant $\kappa > 0$ such that for almost $0 \leq r \leq s \leq t \leq u \leq T$,
    \[
    \E\left[|D_rD_sD_t\sigma_u|^p \right]^{1/p} \leq \kappa(u-t)^{H-\frac{1}{2}}(t-s)^{H-\frac{1}{2}}(s-r)^{H-\frac{1}{2}}.
    \]
\end{hyp}
The result concerning the ATM-IV curvature is the following
\begin{teo} \label{th: main result 2}
    Let $X$ follow model \eqref{Black model}. Assume that the Lévy process $L$ and the volatility process $\sigma$ satisfies Hypotheses \ref{hyp 1} -- \ref{hyp 9}. Define the constants
    \begin{align*}
        C_{vol} = &\frac{\Sigma^2}{4\sigma_0^5(H+\frac{1}{2})^2(2H+2)} \\
        + &\frac{\rho^2\Sigma^2}{4\sigma_0^5(H+\frac{1}{2})^2}\left[ -\frac{6}{(H+\frac{3}{2})^2} + 2 B\left( H+\frac{3}{2}, H + \frac{3}{2}\right) \right] \\
        + & \frac{\rho^2 \Theta}{\sigma_0^3(H+\frac{1}{2})^2}B\left( H+\frac{3}{2}, H + \frac{3}{2}\right),
    \end{align*}
    where $B(\cdot, \cdot)$ denotes the beta function, and
    \[
    C_{jumps} = \frac{(C_+ + C_-)(2-Y)\Gamma\left(1-\frac{Y}{2}\right)}{\sigma_0^{Y+1}Y (Y-1)2^{Y/2}}.
    \]
    Then, the ATM-IV curvature $\partial_{kk}^2 I(T,k^*)$ satisfies the following short-term behaviour
    \begin{align*}
        \lim_{T \to 0} \frac{\partial^2_{kk}I(T,k^*)}{T^{\min\{2H-1, -\frac{Y}{2} \}}} = \begin{cases}
            \displaystyle
            C_{vol} & 2H-1 < -\frac{Y}{2} \\
            \displaystyle
            C_{jumps} & 2H-1 > -\frac{Y}{2} \\
            C_{vol} + C_{jumps} &2H -1 =- \frac{Y}{2}.
        \end{cases}
    \end{align*}
\end{teo}
Theorem \ref{th: main result 2} is proved in Section \ref{sec 4}.

\subsection{Consequences of the main results}
The first implication of the results is that, following Theorem \ref{th: main result}, we have that the leading term in the ATM-IV skew behaves, in a model with rough volatility and infinite variation Lévy jumps, as the most dominant singularity. Even though the result provides a first step in providing a complete negative power expansion of the ATM-IV skew, the result also provides a calibration issue, that is, only with the skew information we are not able to decide from market data whether the dominating behaviour comes from the Lévy process or the rough volatility, which creates a dichotomy in the calibration scheme.

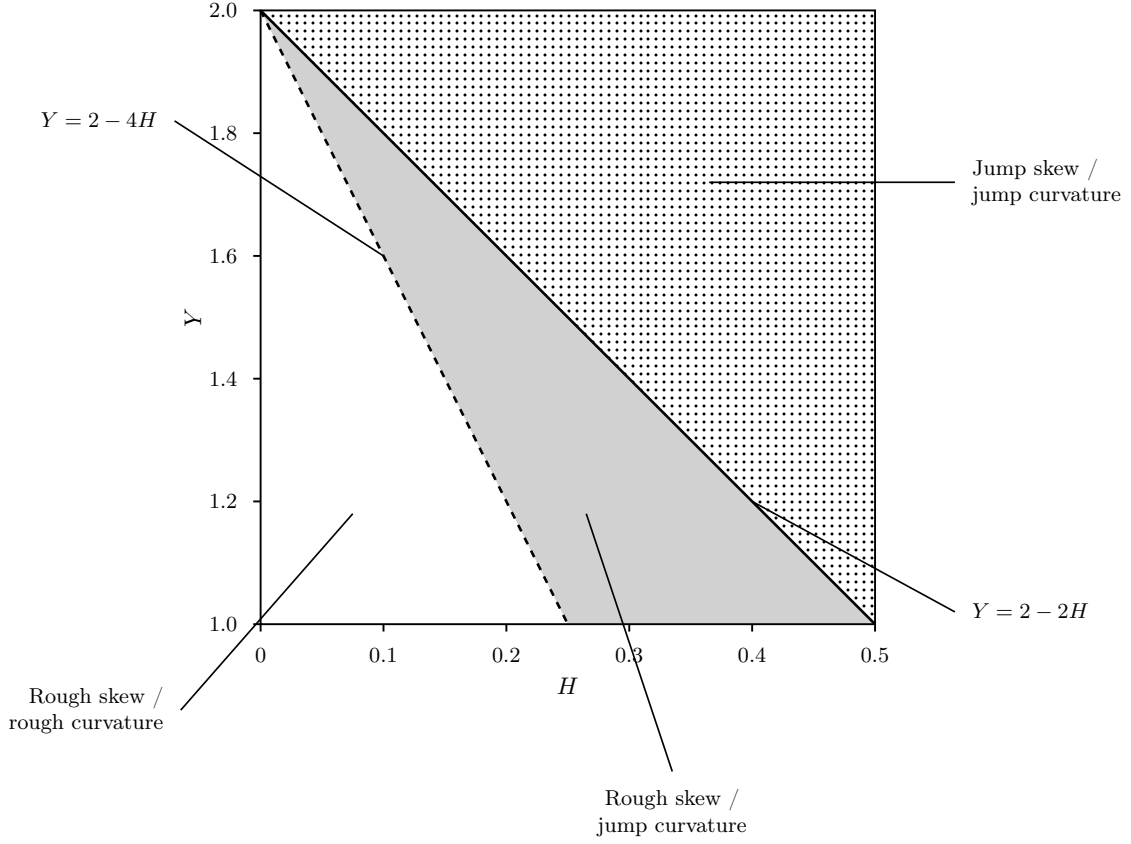
\begin{figure}[H]
    \centering
    \begin{tikzpicture}[
    x=18cm,
    y=9cm,
    every node/.style={font=\small},
    boundary/.style={black, line width=1.1pt},
    axis/.style={black, line width=0.8pt}
]



\fill[white]
    (0,1) -- (0.25,1) -- (0,2) -- cycle;

\fill[black!18]
    (0,2) -- (0.5,1) -- (0.25,1) -- cycle;

\fill[
    fill=white,
    pattern=dots,
    pattern color=black
]
    (0,2) -- (0.5,1) -- (0.5,2) -- cycle;


\draw[boundary]
    (0,2) -- (0.5,1);

\draw[boundary, dashed]
    (0,2) -- (0.25,1);


\draw[axis] (0,1) rectangle (0.5,2);

\foreach \h/\lab in {
    0/{0},
    0.1/{0.1},
    0.2/{0.2},
    0.3/{0.3},
    0.4/{0.4},
    0.5/{0.5}
}{
    \draw[axis] (\h,1) -- ++(0,-0.012);
    \node[anchor=north] at (\h,0.975) {$\lab$};
}

\foreach \yy/\lab in {
    1/{1.0},
    1.2/{1.2},
    1.4/{1.4},
    1.6/{1.6},
    1.8/{1.8},
    2/{2.0}
}{
    \draw[axis] (0,\yy) -- ++(-0.006,0);
    \node[anchor=east] at (-0.012,\yy) {$\lab$};
}

\node[font=\normalsize] at (0.25,0.90) {$H$};
\node[font=\normalsize, rotate=90] at (-0.055,1.5)
    {$Y$};


\draw[black, line width=0.7pt]
    (0.075,1.18) -- (-0.065,0.86);
\node[anchor=east, align=right]
    at (-0.072,0.86)
    {Rough skew /\\rough curvature};

\draw[black, line width=0.7pt]
    (0.265,1.18) -- (0.335,0.76);
\node[anchor=north, align=center]
    at (0.335,0.745)
    {Rough skew /\\jump curvature};

\draw[black, line width=0.7pt]
    (0.365,1.72) -- (0.565,1.72);
\node[anchor=west, align=left]
    at (0.572,1.72)
    {Jump skew /\\jump curvature};


\draw[black, line width=0.7pt]
    (0.10,1.60) -- (-0.070,1.82);
\node[anchor=east, align=right]
    at (-0.077,1.82)
    {$Y=2-4H$};

\draw[black, line width=0.7pt]
    (0.40,1.20) -- (0.565,1.02);
\node[anchor=west, align=left]
    at (0.572,1.02)
    {$Y=2-2H$};

\end{tikzpicture}
    \caption{Dominance regions for the ATM implied-volatility skew and
    curvature in the $(H,Y)$-plane.}
    \label{fig:dominance-regions}
\end{figure}

On the other hand, Theorem \ref{th: main result 2} not only erases the dichotomy but also shows why the combination of rough volatility models together with infinite variation Lévy processes are the most complete answer to explain the short-time behaviour of the implied volatility. Indeed, assume that we decide to model the log-price with a rough volatility process with Hurst index $H$. Theorems \ref{th: main result} and \ref{th: main result 2} show that if the skew and curvature are dominated by terms $\partial_kI(T,k^*) \sim A_sT^{\beta_s}$, $\partial^2_{kk}I(T,k^*) \sim A_cT^{\beta_c}$ then, we have heuristically (because Hypotheses \ref{hyp 1} -- \ref{hyp 9} have to be checked or an approximation argument as in Section \ref{sec 4} has to be performed) that
\[
\beta_c = 2 \beta_s.
\]
Similarly, in log-prices modelled by constant volatility and an infinite variation Lévy process we must have
\[
\beta_c = \beta_s - \frac{1}{2}.
\]
Thus, rough volatility and infinite variation Lévy models, which provide freedom to select $H$ and $Y$ to match the ATM-IV skew, produces a restrictive power-law behaviour in the ATM-IV curvature, thus not providing full flexibility for modelling the implied volatility surface. The solution to get rid of this strict skew-curvature power relationship is by considering a model like model \eqref{Black model} where both the stochastic volatility and the Lévy jumps contribute to the power-law behaviour of the ATM-IV skew and curvature. To represent the skew-curvature power law relationship, we can plot a phase diagram in the variables $(H,Y)$ for $H \in (0,1/2)$ and $Y \in (1,2)$, in which we identify which combination of parameters generate a strict relationship and which combination of parameters yields independent power-law behaviours for the ATM-IV skew and curvature. We represent the regions in Figure \ref{fig:dominance-regions}.

Notice that we also find a meaningful consequence out of the phase diagram in Figure \ref{fig:dominance-regions}, that is, it is not possible for the jumps to dominate the skew and the rough volatility to dominate the curvature at the same time. Indeed, if that was the case, we would have simultaneously
\[
\frac{1-Y}{2} < H- \frac{1}{2}, \quad 2H-1 < -\frac{Y}{2}.
\]
rearranging the variable $Y$ we find that
\[
2-2H < Y < 2-4H,
\]
which is impossible due to the fact that $H > 0$. This discussion can be reformulated as the following statement: If $\beta_s$ and $\beta_c$ denote the leading powers of the ATM-IV skew and curvature respectively and 
\[
\beta_c \neq \beta_s - \frac{1}{2},
\]
then the ATM-IV skew is dominated by the rough volatility part of $X$, solving therefore the non-identifiability of the mechanism generating the ATM-IV skew power law behaviour. Regarding the identification of the mechanism leading the skew and the curvature, the statement has to be taken more carefully. Indeed, we know that if $\beta_c \neq \beta_s -\frac{1}{2}$ and $\beta_c \neq 2\beta_s$ then, we must have a rough dominated skew and a jump dominated curvature. However, it may happen that $\beta_c =2\beta_s$ and the skew and curvature are not rough dominated. This happens in the critical line $Y = 2-4H$.

If the exponent of the curvature lies on the critical line $Y = 2-4H$, our paper alone does not fully give an answer on which mechanism generated the ATM-IV curvature, since the coefficient of the curvature may be either $C_{vol}$ or $C_{vol}+C_{jumps}$. This discussion can be summarized as follows:

Let $\beta_s$ and $\beta_c$ denote the leading-term power of the ATM-IV skew and curvature short-term expansions. Then:
\begin{enumerate}
    \item If $\beta_c = \beta_s - \frac{1}{2}$ then we have:
    \begin{itemize}
        \item Skew dominated by Lévy (or mixed on the critical line $Y = 2-2H$) dynamics.
        \item Curvature dominated by Lévy dynamics.
    \end{itemize}
    \item If $\beta_c = 2\beta_s$ then we have:
    \begin{itemize}
        \item Skew dominated by rough volatility dynamics.
        \item Curvature dominated by rough (or mixed on the critical line $Y = 4-2H$) dynamics.
    \end{itemize}
    \item If $\beta_s - \frac{1}{2}< \beta_c < 2\beta_s$ then we have:
    \begin{itemize}
        \item Skew dominated by rough volatility dynamics.
        \item Curvature dominated by Lévy dynamics.
    \end{itemize}
\end{enumerate}

\section{The Implied Volatility Skew} \label{sec 3}
From now on, we will use the following notation throughout the paper:
\begin{itemize}
    \item We define $v_t := \sqrt{\frac{1}{T-t}\int_t^T \sigma_s^2 ds}$.
    \item We define $G(T,t,x, k,\sigma) := (\partial_{xx}^2 - \partial_x)\BS(T,t,x,k,\sigma)$. We drop the variable $k$ when we evaluate at $k = k^*$.
\end{itemize}

As mentioned previously, in \cite{ALV} the authors provide an analysis for $\partial_{X_0}I(T,k^*)$ for the case where the Lévy process is a compound Poisson process. Since we want to reuse as much of that analysis as possible, we will analyse $\partial_{X_0}I(T,k^*)$ for our model and we will deduce Theorem \ref{th: main result} by applying the conversion $\partial_{X_0}I(T,k^*) = -\partial_kI(T,k^*)$. In \cite{ALV} the authors prove an expansion that can be adapted to our context thanks to the results in \cite{jafari2013hull}:
\begin{align*}
    \partial_{\sigma}\BS(T,0,X_0,I(T,k^*))\partial_{X_0}I(T,k^*) = & \frac{\rho}{2}\E\left[\int_0^T (\partial_x - \frac{1}{2}) \partial_x G(T,s,X_s,v_s)\Lambda_s ds \right] \\
    + &\E\left[\int_0^T \int_{\R} \Delta_{yy} (\partial_x - \frac{1}{2})\BS(T,s,X_s,v_s) \nu(dy) ds\right] \\
    - & c_2 \E\left[\int_0^T (\partial_x - \frac{1}{2}) \partial_x \BS(T,s,X_s,v_s) ds \right] \\
    =: &I_1(T) + I_2(T) + I_3(T).
\end{align*}
Here,
\[
\Lambda_s = \sigma_s \int_s^T D_s\sigma_r^2 dr ,
\]
\[
\Delta_{yy}f(x) = f(x+y) - f(x) - y \partial_xf(x),
\]
and
\[
c_2 = \int_{\R} (e^y-1-y) \nu(dy).
\]
The analysis of the blow-up effect of the ATM-IV skew for our model relies on the analysis of the terms $I_1(T)$, $I_2(T)$ and $I_3(T)$. The following auxiliary results are key in order to properly perform an analysis on the previous integrals. The first lemma is an already well-known tool used in the Malliavin calculus approach for the study of the implied volatility surface.
\begin{lema} \label{classic bound greeks}
    Let $0 \leq t \leq s < T$ and $\mathcal{G}_t:= \mathcal{F}_t \vee \mathcal{F}^{W}_T$. Then, for every $n \geq 0$, there exists $C = C(n,\rho) > 0$ such that
    \[
    \left|E\left( \partial^n_x G(T,s,X_s,k,v_s) | \mathcal{G}_t\right) \right|\leq C \left( \int_t^T \sigma_s^2 ds\right)^{-\frac{n+1}{2}},
    \]
    where $G = (\partial^2_{xx} - \partial_x) \BS$.
\end{lema}
\begin{proof}
    See \cite{ALV}.
\end{proof}
The following two lemmas aim to study the limit distribution of $X_t$ defined as in \eqref{Black model} when $t \to 0$.
\begin{lema} \label{almost constant}
    Let $X$ be defined as in \eqref{Black model}. Then,
    \[
    X_t = X_0 + \sigma_0 (\rho W_t + \sqrt{1-\rho^2}B_t) + R_t
    \]
    where $R_t$ satisfies
    \[
    \lim_{t \to 0} \frac{R_t}{\sqrt{t}} = 0
    \]
    in distribution.
\end{lema}
\begin{proof}
    We can write $X_t = X_0 + \sigma_0 Z_t + R_t$, where
    \[
    Z_t = \rho W_t + \sqrt{1-\rho^2}dB_t,
    \]
    and
    \[
    R_t = -\eta t +\int_0^t (\sigma_s-\sigma_0)(\rho dW_s + \sqrt{1-\rho^2}dB_s) - \frac{1}{2}\int_0^t \sigma_s^2 ds + L_t = R^{(1)}_t + L_t.
    \]
    By the continuity of the sample paths of $\sigma$ we have
    \[
    \lim_{t \to 0}\frac{1}{t}\int_0^t (\sigma_s - \sigma_0)^2 ds = 0
    \]
    almost surely. Hence,
    \[
    \lim_{t \to 0} \frac{\left(\int_0^t \sigma_s^2 ds \right)^2}{t} = \lim_{t \to 0} \frac{\left(\int_0^t \sigma_s^2 ds \right)}{t}\int_0^t \sigma_s^2 ds = 0.
    \]
   The Itô isometry shows that
   \[
   \lim_{t \to 0} \frac{\int_0^t (\sigma_s- \sigma_0)(\rho dW_s + \sqrt{1-\rho^2}dB_s)}{\sqrt{t}} = 0
   \]
   in probability. This, together with the fact that
   \[
   \lim_{t \to 0} \frac{\eta t}{\sqrt{t}} = 0, \quad \lim_{t \to 0} \frac{\int_0^t \sigma_s^2 ds}{\sqrt{t}} = 0
   \]
   shows that $t^{-1/2} R_t^{(1)} \to 0$ in distribution. We finally want to prove the weakly convergence of $t^{-1/2}L_t$ towards $0$. The characteristic function of $L_t$ is
    \[
    \E\left[e^{iuL_t} \right] = e^{t\psi(t)},  \quad \psi(t) = \int_{\R}(e^{iuy}-1-iuy\1_{\{|y|\leq 1\}})\nu(dy).
    \]
    The characteristic function of $t^{-1/2} L_t$ is
    \[
     \E\left[e^{iu t^{-1/2}L_t} \right] = e^{t\psi(ut^{-1/2})}
    \]
    We will show that this quantity converges to $1$ for all $u \in \R$. Observe that
    \[
    t\psi(ut^{-1/2}) = t\int_{\R} (e^{iut^{-1/2}y}-1-iut^{-1/2}y\1_{\{|y|\leq 1\}})\nu(dy) = A(t) + B(t),
    \]
    where
    \[
    A(t) = t\int_{|y| > 1} (e^{iut^{-1/2}y}-1)\nu(dy), \quad B(t) =t\int_{|y| \leq 1} (e^{iut^{-1/2}y}-1-iut^{-1/2}y)\nu(dy).
    \]
    Regarding $A(t)$, notice that
    \[
    |A(t)|\leq t \int_{|y| > 1} |e^{iut^{-1/2}y} - 1|\nu(dy) \leq 2t\nu(\{|y| > 1\}) \to 0
    \]
    as $t \to 0$. Regarding $B(t)$, choose $\beta \in (Y,2)$. Then, since $q(y) \to 1$ as $y \to 0$ and $q$ is continuous on each semi-axis, there exists a constant $M$ such that
    \[
    |q(y)| \leq M, \quad 0 < |y| \leq 1.
    \]
    Then, using the bound
    \[
    |e^{ix}-1-ix| \leq C_{\beta} |x|^{\beta}
    \]
    we find that
    \[
    \left| t\int_{|y| \leq 1} (e^{iut^{-1/2}y}-1-iut^{-1/2}y)\nu(dy)\right| \leq C_{\beta} |u|^{\beta} t^{1- \frac{\beta}{2}} \int_{|y| \leq 1} |y|^{\beta} \nu(dy).
    \]
    This term goes to zero as $t \to 0$ because
    \[
    \int_{|y| \leq 1}|y|^{\beta} \nu(dy)  \leq M(C_+ + C_{-}) \int_0^1 y^{\beta-1-Y}\nu(dy) < \infty.
    \]
    Hence, since $A(t) + B(t) \to 0$ we conclude that $t^{-1/2} L_t \to 0$ in distribution, concluding therefore the desired result.
\end{proof}
\begin{lema}\label{CLT}
    Let $\lambda \in (0,1)$ and let $u_t = v_t \sqrt{T-t}$. Then
    \[
    \lim_{T \to 0} \frac{X_{\lambda T} - X_0}{u_{\lambda T}} = Z_{\lambda} \sim N\left(0, \frac{\lambda}{1-\lambda}\right),
    \]
    where the limit is taken in distribution.
\end{lema}
\begin{proof}
    We write
    \[
    X_{\lambda T} -X_0 = \sigma_0 Z_{\lambda T} + R_{\lambda T}.
    \]
    From the previous lemma, we see that $T^{-1/2} R_{\lambda T} \to 0$ in distribution. Now, since
    \[
    u_{\lambda T} = \sqrt{\int_{\lambda T}^T \sigma_s^2 ds},
    \]
    we have that
    \[
    \lim_{T \to 0}\frac{u_{\lambda T}^2}{T} = \lim_{T \to 0} \frac{1}{T}\int_{\lambda T}^T \sigma_s^2ds = (1-\lambda)\sigma_0^2
    \]
    where the previous limit is taken almost surely. We write
    \begin{align*}
        \frac{X_{\lambda T} -X_0}{u_{\lambda T}} = &\frac{\sigma_0 W_{\lambda T} + R_{\lambda T}}{u_{\lambda T}} \cdot \frac{\sqrt{T}}{\sqrt{T}}  =\left( \frac{\sigma_0W_{\lambda T}}{\sqrt{T}} + \frac{R_{\lambda T}}{\sqrt{T}}\right)\frac{\sqrt{T}}{u_{\lambda T}}.
    \end{align*}
    Using that
    \[
   \frac{\sigma_0W_{\lambda T}}{\sqrt{T}} \to N\left(0, \sigma_0^2 \lambda \right), \quad \frac{\sqrt{T}}{u_{\lambda T}} \to \frac{1}{\sigma_0 \sqrt{1-\lambda}}, \quad \frac{R_{\lambda T}}{\sqrt{T}} \to 0
    \]
    we have, by Slutski's theorem, that
    \[
    \lim_{T \to 0} \frac{X_{\lambda T} - X_0}{u_{\lambda T}} = Z_{\lambda} \sim N\left(0, \frac{\lambda}{1-\lambda}\right)
    \]
    as desired.
\end{proof}
We have now enough tools to analyse the three terms $I_1(T)$, $I_2(T)$ and $I_3(T)$.
\subsection{Analysis of the integrals}
With the previous results we are able to understand the asymptotic behaviour of $I_1(T)$, $I_2(T)$ and $I_3(T)$. In particular, we will show that $I_1(T)$ yields the contribution of the stochastic volatility term, $I_2(T)$ yields the contribution of the jumps and $I_3(T)$ is negligible with respect to $I_1(T)$ and $I_2(T)$.
\begin{prop} \label{prop: I3}
    We have $I_3(T) = O(\sqrt{T})$.
\end{prop}
\begin{proof}
    Observe that
    \[
    \left(\partial_x - \frac{1}{2}\right) \partial_x\BS(T,s,X_s,v_s) =G(T,s,X_s,v_s) + \frac{1}{2}\partial_x \BS(T,s,X_s,v_s).
    \]
    The argument is the same as in \cite{ALV}. Indeed, we use that $ |\partial_x \BS(T,s,X_s,v_s)| \leq e^{X_s} + K$ and Lemma \ref{classic bound greeks} to find the bound.
\end{proof}
Notice that we are not interested in the limit of $T^{-1/2}I_3(T)$ because, as we will see in the following propositions, $I_3(T)$ is negligible compared to $I_1(T)$ and $I_2(T)$.
\begin{prop} \label{prop: I2}
    We have 
    \[
    \lim_{T \to 0}\frac{I_2(T)}{T^{1-Y/2}} = e^{X_0}\sigma_0^{-Y}(C_{-} - C_{+})\frac{\Gamma\left(\frac{3-Y}{2} \right)}{Y(Y-1)2^{Y/2}\sqrt{\pi}}.
    \]

\end{prop}
\begin{proof}
    In order to streamline the notation, let $u_s = v_s \sqrt{T-s}$ and let $\mathcal{B}(s,X_s,v_s) = (\partial_x - \frac{1}{2})\BS(T,s,X_s,v_s)$.
    Notice that
    \[
    \mathcal{B}(s,X_s,v_s) = \frac{1}{2}\left(e^{X_s} \Phi(d_+) + e^{X_0}\Phi(d_{-}) \right).
    \]
    We write
    \[
    I_2(T) = I_2(T;\delta) + I_2^c(T;\delta),
    \]
    where $I_2(T;\delta)$ corresponds to the integral over the set $\{|y| \leq \delta\}$. Regarding $I_2^c(T;\delta)$ we have, using the explicit expression for $\mathcal{B}$ and $\partial_x \mathcal{B}$,
    \[
    I_2^c(T;\delta) \leq \int_0^T \int_{|y| > \delta} \left[ e^{X_0}\left(\frac{1}{2}e^y  + \frac{3}{2}+\frac{|y|}{2}\right) + \frac{|y|e^{X_0}}{\sqrt{2\pi}}\E\left[\frac{1}{u_s} \right]\right]ds \nu(dy).
    \]
    Using the Hypotheses on $\sigma$ we find that
    \[
    |I_2^c(T; \delta)| \leq C_{\delta,\nu}(T + \sqrt{T}),
    \]
    showing that
    \begin{equation} \label{eq: 2 term vanishes}
    \lim_{T \to 0}T^{Y/2 - 1}I_2^c(T;\delta) = 0.
    \end{equation}
    We now consider the term
    \[
    I_2(T;\delta) = \E \left[\int_0^T \int_{|y|\leq \delta} \Delta_{yy} \B(s,X_s,v_s) \nu(dy) ds\right].
    \]
    Recall that $u_s = v_s\sqrt{T-s}$. For $s = \lambda T$ we define
    \[
    A_{\lambda T} := \frac{u_{\lambda T}}{\sqrt{T}}.
    \]
    The change of variables of integration $s = \lambda T$ and $y = u_{\lambda T} z$ on $I_2(T;\delta)$ allow us to write
    \[
    I_2(T;\delta) = T^{1 - \frac{Y}{2}}\E\left[\int_0^1 A_{\lambda, T}^{-Y}J(\lambda,T;\delta) d\lambda \right],
    \]
    where
    \[
    J(\lambda, T; \delta) := \int_{\R} H_{\lambda,T}(z) C\left( \frac{z}{|z|}\right) q(u_{\lambda T}z) \1_{\{|u_{\lambda T}z| \leq  \delta\}}|z|^{-1-Y }dz
    \]
    and
    \begin{align*}
    H_{\lambda, T}(z) = &\B(\lambda T, X_{\lambda T} + u_{\lambda T}z, v_{\lambda T}) - \B(\lambda T, X_{\lambda T}, v_{\lambda T})  \\
    -& u_{\lambda T}z\partial_x \B(\lambda T, X_{\lambda T}, v_{\lambda T}).
    \end{align*}
    By computing explicitely $\B$ and $\partial_x \B$ we can show that, if we define
    \[
    \xi_{\lambda T} := \frac{X_{\lambda T}-X_0}{u_{\lambda T}},
    \]
    the term $H_{\lambda, T}$ reads as
    \begin{align*}
        H_{\lambda,T}(z) = & \frac{1}{2}e^{X_{\lambda T} + u_{\lambda T}z}\Phi\left( \xi_{\lambda T} +z+ \frac{u_{\lambda T}}{2}\right) + \frac{1}{2}e^{X_0}\Phi \left( \xi_{\lambda T} + z - \frac{u_{\lambda T}}{2}\right) \\
        - & \frac{1}{2} e^{X_{\lambda T}}\Phi\left( \xi_{\lambda T} + \frac{u_{\lambda T}}{2}\right) - \frac{1}{2}e^{X_0}\Phi \left( \xi_{\lambda T} - \frac{u_{\lambda T}}{2}\right) \\
        - &\frac{1}{2}u_{\lambda T}ze^{X_{\lambda T}}\Phi\left(\xi_{\lambda T}+ \frac{u_{\lambda T}}{2} \right) - ze^{X_0} \phi\left( \xi_{\lambda T} - \frac{u_{\lambda T}}{2}\right).
    \end{align*}
    The continuity of the paths of $\sigma$ allow us to write
\[
\lim_{T \to 0} A_{\lambda, T} = \sqrt{1-\lambda}\sigma_0 \quad \text{ a.s.},
\]
and therefore the convergence also holds in distribution. Moreover, by Lemma \ref{CLT} we have that
\[
\xi_{\lambda T} \to Z_{\lambda} \sim N\left(0, \frac{\lambda}{1-\lambda} \right).
\]
Since $X_{\lambda T} \to X_0$ and $u_{\lambda T} \to 0$ in probability, Slutsky's theorem ensures that, for every fixed $z \in \R$, 
\[
\lim_{T \to 0} H_{\lambda, T}(z) = \mathcal{L}(Z_{\lambda},z) :=  e^{X_0}\left( \Phi(Z_{\lambda} + z) - \Phi(Z_{\lambda}) - z \phi(Z_{\lambda})\right).
\]
Observe that, since $u_{\lambda T} \to 0$, then $q(u_{\lambda T} z) \to 1$ and, moreover, the indicator $\1_{\{|u_{\lambda T} z| \leq \delta\}} \to  1$. Hence, the dominated convergence theorem allows us to write
\[
\lim_{T \to 0} \frac{I_2(T;\delta)}{T^{1-\frac{Y}{2}}} = \int_0^1 \sigma_0^{-Y}(1-\lambda)^{-Y/2} \int_{\R} \E[\mathcal{L}(Z_{\lambda},z)]\nu'(dz) d\lambda,
\]
where
\[
\nu'(dz) = \frac{1}{|z|^{1+Y}}(C_+\1_{\{z>0\}} + C_{-} \1_{\{z<0\}})dz.
\]
The last step before concluding our result is the computation of the term $ \E[\mathcal{L}(Z_{\lambda},z)]$. In order to streamline the notation, we write
\[
\Lambda:= \frac{\lambda}{1-\lambda}.
\]
Then, since $Z_{\lambda} \sim N(0,\Lambda)$, it requires elementary computations to check that
\[
\E[\Phi(Z_{\lambda}+z)]= \Phi\left(\frac{z}{\sqrt{1+\Lambda}} \right), \quad \E[\Phi(Z_{\lambda})] = \frac{1}{2},
\]
and
\[
\E\left[\phi(Z_{\lambda}) \right] = \frac{1}{\sqrt{2\pi(1+\Lambda)}}.
\]
Thus,
\[
\int_{\R} \E[\mathcal{L}(Z_{\lambda},z)] \nu'(dz) = e^{X_0}\int_{\R} \left( \Phi\left(\frac{z}{\sqrt{1+\Lambda}} \right) -\frac{1}{2} - \frac{z}{\sqrt{2\pi(1+\Lambda)}}\right)\nu'(dz).
\]
The change of variables $w = z(1+\Lambda)^{-1/2}$ allows us to write
\[
\int_{\R} \E[\mathcal{L}(Z_{\lambda},z)] \nu'(dz) = e^{X_0}(1+\Lambda)^{-Y/2}\int_{\R} (\Phi(w) - \frac{1}{2}- \frac{w}{\sqrt{2\pi}}) \nu'(dw).\]
Let
\[
f(w) = \Phi(w) - \frac{1}{2}- \frac{w}{\sqrt{2\pi}}.
\]
Since $f$ is odd, we can write
\[
\int_{\R} f(w)\nu'(dw) = (C_+- C_{-}) \int_0^{\infty} f(w) w^{-1-Y}dw.
\]
Integrating by parts,
\[
\int_0^{\infty} f(w) w^{-1-Y}dw = \frac{1}{Y\sqrt{2\pi}}\int_0^{\infty} \left( e^{-\frac{w^2}{2}} - 1\right)w^{-Y}dw.
\]
Integrating by parts another time allows us to conclude that
\[
\int_{\R} f(w) \nu'(dw) = (C_{-}-C_{+}) \frac{\Gamma\left( \frac{3-Y}{2}\right)}{Y(Y-1)2^{Y/2}\sqrt{\pi}}.
\]
Since $1+\Lambda = \frac{1}{1-\lambda}$ we have
\[
\int_{\R} \E[\mathcal{L}(Z_{\lambda},z)]\nu'(dz) = e^{X_0}(C_- - C_+)\frac{\Gamma\left(\frac{3-Y}{2}\right)}{Y(Y-1)2^{Y/2}\sqrt{\pi}}(1-\lambda)^{Y/2}.
\]
Therefore,
\[
\lim_{T \to 0} \frac{I_2(T;\delta)}{T^{1-\frac{Y}{2}}} = e^{X_0}\sigma_0^{-Y}(C_{-} - C_+) \frac{\Gamma\left( \frac{3-Y}{2}\right)}{Y(Y-1)2^{Y/2}\sqrt{\pi}}.
\]
Since
\[
\lim_{T \to 0} \frac{I_2(T)}{T^{1-\frac{Y}{2}}} = \lim_{T \to 0} \frac{I_2(T;\delta)}{T^{1-\frac{Y}{2}}} + \lim_{T \to 0} \frac{I^c_2(T;\delta)}{T^{1-\frac{Y}{2}}},
\]
we use Equation \eqref{eq: 2 term vanishes} to find that
\[
\lim_{T \to 0} \frac{I_2(T)}{T^{1-\frac{Y}{2}}} = e^{X_0}\sigma_0^{-Y}(C_{-} - C_+) \frac{\Gamma\left( \frac{3-Y}{2}\right)}{Y(Y-1)2^{Y/2}\sqrt{\pi}},
\]
as desired.
\end{proof}
Finally, we deal with the asymptotic behaviour of $I_1(T)$.
\begin{prop} \label{prop: I1}
    We have
\[
\lim_{T\to0}\frac{I_1(T)}{T^H}
=
 -\frac{\rho e^{X_0}\Sigma}{2\sqrt{2\pi}\sigma_0^2 \left(H+\frac{1}{2} \right)\left(H+\frac{3}{2} \right)}.
\]
\end{prop}

\begin{proof}
    Recall that
    \[
    I_1(T) = \frac{\rho}{2}\E\left[ \int_0^T L(s,X_s,v_s)\Lambda_s ds\right],
    \]
    where
    \[
    L(s,X_s,v_s) =  \frac{e^{X_s}\phi(d_+)}{2u_s^3}\left(u_s^2 - 3d_+u_s + 2d_+^2 -2 \right),
    \]
    and
    \[
    \Lambda_s = \sigma_s \int_s^T D_s \sigma_r^2 dr.
    \]
    For $\lambda \in [0,1]$ we define the quantities
    \[
    A_{\lambda T} = \frac{u_{\lambda T}}{\sqrt{T}}, \quad B_{\lambda T} = \frac{\Lambda_{\lambda T}}{T^{H+\frac{1}{2}}},
    \]
    and
    \[
    \xi_{\lambda T} = \frac{X_{\lambda T}-X_0}{u_{\lambda T}}.
    \]
    By means of the change of variables $s = \lambda T$ we have
    \[
    \frac{I_1(T)}{T^H} = \frac{\rho}{4}\int_0^1 \E\left[e^{X_{\lambda T}}A_{\lambda T}^{-3}B_{\lambda T}\phi(d_+)\left(TA^2_{\lambda T} -3\sqrt{T}A_{\lambda T}d_+ +2d_+^2 -2 \right) \right]d\lambda.
    \]
    Again, the continuity of the sample paths of $\sigma$ show that
    \[
    \lim_{T \to 0}A_{\lambda T} = \sigma_0 \sqrt{1-\lambda} 
    \]
    almost surely. We now can write $B_{\lambda T}$ as
    \[
    B_{\lambda T} = \sigma_{\lambda T} \int_{\lambda}^1 D_{\lambda T}\sigma_{Tu}^2 \frac{(u-\lambda)^{H-\frac{1}{2}}}{(T(u-\lambda))^{H-\frac{1}{2}}} du.
    \]
    Hence, Hypotheses \ref{hyp 4} -- \ref{hyp 6} show that for a fixed $0<\lambda < 1$ we have
    \[
    B_{\lambda T} \to \sigma_0 \Sigma \int_{\lambda}^1 (u-\lambda)^{H-\frac{1}{2}}du = \frac{\sigma_0 \Sigma}{H+\frac{1}{2}}(1-\lambda)^{H+\frac{1}{2}}.
    \]
    where the convergence holds in $L^1(\Omega)$. Lemma \ref{CLT} together with Slutsky's theorem shows that 
    \[
    (X_{\lambda T}, A_{\lambda T}, B_{\lambda T}, d_1) \to \left(X_0, \sigma_0\sqrt{1-\lambda},  \frac{\sigma_0 \Sigma}{H+\frac{1}{2}}(1-\lambda)^{H+\frac{1}{2}}, Z_{\lambda}\right).
    \]
    Thus, applying the dominated convergence theorem we can show that
    \[
    \lim_{T \to 0}\frac{I_1(T)}{T^H} = \frac{\rho  e^{X_0}\Sigma}{2\sigma_0^2(H+\frac{1}{2})} \int_0^1 (1-\lambda)^{H-1}\E[\phi(Z_{\lambda})(Z_{\lambda}^2-1)]d\lambda.
    \]
    Again, let $\Lambda = \frac{\lambda}{1-\lambda}$ so that $Z_{\lambda} \sim N(0,\Lambda)$. Then,
    \[
    \E[\phi(Z_{\lambda})] = \frac{1}{\sqrt{2\pi(1+\Lambda)}}, \quad \E[Z_{\lambda}^2 \phi(Z_{\lambda})] = \frac{\Lambda}{\sqrt{2\pi}(1+\Lambda)^{3/2}}
    \]
    and, therefore,
    \[
    \E[\phi(Z_{\lambda})(Z_{\lambda}^2 -1)] =-\frac{1}{\sqrt{2\pi}(1+\Lambda)^{3/2}}.
    \]
    Finally, using that $1+\Lambda = (1-\lambda)^{-1}$ we conclude that
    \[
    \lim_{T \to 0} \frac{I_1(T)}{T^{H}} = -\frac{\rho e^{X_0}\Sigma}{2\sqrt{2\pi}\sigma_0^2 \left(H+\frac{1}{2} \right)\left(H+\frac{3}{2} \right)},
    \]
    getting the desired conclusion.
\end{proof}
We are now ready to prove our main result.
\begin{proof}(Theorem \ref{th: main result}). The relationship $\partial_kI(T,k^*) = -\partial_{X_0}I(T,k^*)$ allows us to write
\[
\partial_kI(T,k^*) = -\frac{I_1(T)+I_2(T)+I_3(T)}{\partial_{\sigma}\BS(T,0,X_0, I(T,k^*))}.
\]
If we show that $I(T,k^*) \to  \sigma_0$ as $T \to 0$ then we will conclude the result by the elementary computation:
\[
\lim_{T \to 0} \frac{\partial_{\sigma}\BS(T,0,X_0, \sigma_0)}{\sqrt{T}} = \frac{e^{X_0}}{\sqrt{2\pi}}.
\]
We start from the identity
\[
\mathcal{C}:= \mathcal{C}(T,k^*) = e^{X_0} \left( 2\Phi\left( \frac{I(T,k^*) \sqrt{T}}{2}\right)-1 \right).
\]
Manipulating the right-hand side, we obtain
\[
\frac{\mathcal{C}}{\sqrt{T}} = e^{X_0}I(T,k^*) \frac{ \left( 2\Phi\left( \frac{I(T,k^*) \sqrt{T}}{2}\right)-1 \right)}{I(T,k^*) \sqrt{T}}.
\]
A Taylor expansion shows that
\[
e^{X_0}I(T,k^*) \frac{ \left( 2\Phi\left( \frac{I(T,k^*) \sqrt{T}}{2}\right)-1 \right)}{I(T,k^*) \sqrt{T}} = \frac{e^{X_0}I(T,k^*)}{\sqrt{2\pi}} + o(1).
\]
From \cite{JBV2025} we know that the short-time behaviour of the at-the-money call option $\mathcal{C}$ is
\[
\lim_{T \to 0}\frac{\mathcal{C}}{\sqrt{T}} = \frac{e^{X_0}\sigma_0}{\sqrt{2\pi}}.
\]
Hence, the expansion for $\partial_kI(T,k^*)$ follows from applying Propositions \ref{prop: I3}, \ref{prop: I2} and \ref{prop: I1}, adjusting the coefficients obtained in the propositions dividing them by $\frac{e^{X_0}}{\sqrt{2\pi}}$ and reducing the order in $T$ of the integrals by a factor of $\sqrt{T}$.
\end{proof}
This section concludes the proof of Theorem \ref{th: main result}. In the following section we deal with the proof of Theorem \ref{th: main result 2}.
\section{The Implied Volatility Curvature} \label{sec 4}
The tools required for the study of the ATM-IV curvature are essentially the same as the ones for the ATM-IV skew. Therefore, the novelty of this section relies on the result rather than on the methodology. First, we show a suitable decomposition formula for the curvature for the sake of completeness.
\begin{prop}
    Let $X$ follow model \eqref{Black model} and let $I(T,k^*)$ denote the ATM-IV. Then,
    \begin{align*}
    \partial_{\sigma}\BS(T,0,X_0,I(T,k^*))\partial^2_{X_0 X_0}I(T,k^*) =  & \E[G(T,0,X_0,v_0)]-G(T,0,X_0, I(T,k^*)) \\
    + & \frac{\rho}{2}\E\left[ \int_0^T (\partial^2_{xx}-\frac{1}{2})\partial_xG(T,s,X_s,v_s)\Lambda_s ds\right] \\
    + & \E\left[\int_0^T \Delta_{yy}(\partial_{xx}^2-\frac{1}{2}) \BS(T,s,X_s,v_s) \nu(dy) ds \right] \\
    -&c_2 \E\left[ \int_0^T (\partial_{xx}^2-\frac{1}{2}) \partial_x \BS(T,s,X_s,v_s) ds\right] \\
    -&2\partial_{x\sigma}^2\BS(T,0,X_0,I(T,k^*))\partial_{X_0}I(T,k^*) \\
    - & \partial_{\sigma}^2 \BS(T,0,X_0,I(T,k^*)) (\partial_{X_0}I(T,k^*))^2.
\end{align*}
\end{prop}
\begin{proof}
    We start from the formula from \cite{jafari2013hull} for the Call price with log-strike $k$, that is:
    \begin{eqnarray}
        \mathcal{C}(T,k) &= &\E[\BS(T,0,X_0,v_0)] \notag \\
       & + &\frac{\rho}{2}\E\left[ \int_0^T  \partial_x G(T,s,X_s,k,v_s) \Lambda_s ds\right]\notag \\
       & - & c_2 \E\left[\int_0^T \partial_x \BS(T,s,X_s,k,v_s) ds\right]\notag\\
       & + & \E\left[\int_0^T \int_{\R} \Delta_{yy} \BS(T,s,X_{s-},k,v_s) \nu(dy) ds \right].  \label{JafariVives}
    \end{eqnarray}
    Using that $\mathcal{C}(T,k) = \BS(T,0,X_0,k,I(T,k))$ and differentiating with respect to $x$ we find 
    \begin{align*}
        \partial_{\sigma}\BS(T,0,X_0,k,I(T,k)) \partial_{X_0}I(T,k) = &\E[\partial_x \BS(T,0,X_0,k,v_0)] - \partial_x\BS(T,0,X_0,k,I(T,k)) \\
        + & \frac{\rho}{2}\E\left[\int_0^T \partial^2_{xx} G(T,s,X_s,k,v_s) \Lambda_s ds \right] \\
        - & c_2\E\left[\int_0^T \partial^2_{xx} \BS(T,s,X_s,k,_s) ds \right] \\
        + & \E\left[\int_0^T \int_{\R} \Delta_{yy} \partial_x\BS(T,s,X_{s-},k,v_s)\nu(dy)ds \right].
    \end{align*}
   Taking again derivatives with respect to $x$ yields
    \begin{align*}
         \partial_{\sigma}\BS(T,0,X_0,k,I(T,k)) \partial^2_{X_0 X_0}I(T,k) =&\E[\partial^2_{xx} \BS(T,0,X_0,k,v_0)] - \partial^2_{xx}\BS(T,0,X_0,k,I(T,k)) \\
         + & \frac{\rho}{2}\E\left[\int_0^T \partial^3_{xx} G(T,s,X_s,k,v_s) \Lambda_s ds \right] \\
          - & c_2\E\left[\int_0^T \partial^3_{xx} \BS(T,s,X_s,k,_s) ds \right] \\
          + & \E\left[\int_0^T \int_{\R} \Delta_{yy} \partial^2_{xx}\BS(T,s,X_{s-},k,v_s)\nu(dy)ds \right] \\
           -&2\partial_{x\sigma}^2\BS(T,0,X_0,I(T,k^*))\partial_{X_0}I(T,k^*) \\
    - & \partial_{\sigma}^2 \BS(T,0,X_0,I(T,k^*)) (\partial_{X_0}I(T,k^*))^2.
    \end{align*}
    We now evaluate at $k = k^*$. We use that $G = (\partial_{xx}^2-\partial_x)\BS$ and the fact that $\partial_x \BS(T,0,X_0,v_0) = \frac{1}{2}\BS(T,0,X_0,v_0) + \frac{1}{2}e^{X_0}$,  so that we write
    \begin{align*}
    \E[\partial^2_{xx} \BS(T,0,X_0,v_0)] - \partial^2_{xx}\BS(T,0,X_0,I(T,k^*)) = &\E[G(T,0,X_0,v_0)] - G(T,0,X_0,I(T,k^*)) \\
    +& \frac{1}{2}\left( \BS(T,0,X_0,v_0) - \E\left[\BS(T,0,X_0,v_0)\right]\right). \\
    \end{align*}
    The term $ \BS(T,0,X_0,v_0) - \E\left[\BS(T,0,X_0,v_0)\right]$ is substituted by Equation \eqref{JafariVives} and the desired formula is obtained by joining the similar terms.
\end{proof}
Observe that, since $\partial^2_{X_0 X_0}I(T,k^*) = \partial^2_{kk}I(T,k^*)$, the analysis of the terms on the right-hand side yields the analysis of the ATM-IV curvature. For $X$ following model \eqref{Black model}, we write
\[
X_t = X_t^{(c)} + L_t - \eta t
\]
so that
\[
X_t^{(c)} = X_0+ \int_0^t \sigma_s(\rho dW_s + \sqrt{1-\rho^2}dB_s) - \frac{1}{2}\int_0^t \sigma_s^2 ds.
\]
The analysis of the curvature results in the analysis of the following terms:
\begin{align*}
\tilde{I}_1(T) =&\E[G(T,0,X_0,v_0)] -G(T,0,X_0, I^{(c)}(T,k^*))\\
+ &\frac{\rho}{2}\E\left[ \int_0^T(\partial^2_{xx}-\frac{1}{2})\partial_xG(T,s,X^{(c)}_s,v_s)\Lambda_s ds\right],
\end{align*}
\[
\tilde{I}_2(T) = G(T,0,X_0, I^{(c)}(T,k^*)) - G(T,0,X_0, I(T,k^*)),
\]
\[
\tilde{I}_3(T) = \frac{\rho}{2}\E\left[\int_0^T (\partial^2_{xx}-\frac{1}{2})\partial_x(G(T,s,X_s,v_s)  - G(T,s,X^{(c)}_s,v_s))\Lambda_s ds \right],
\]
\[
\tilde{I}_4(T) = \E\left[ \int_0^T \Delta_{yy} (\partial^2_{xx}-\frac{1}{2}) \BS(T,s,X_s,v_s)\nu(dy) ds \right],
\]
\[
\tilde{I}_5(T) = -c_2 \E\left[ \int_0^T (\partial^2_{xx}-\frac{1}{2}) \partial_x \BS(T,s,X_s,v_s) ds\right],
\]
\[
\tilde{I}_6(T) = -2\partial_{x\sigma}^2\BS(T,0,X_0,I(T,k^*))\partial_{X_0}I(T,k^*) 
\]
and 
\[
\tilde{I}_7(T) = - \partial_{\sigma}^2 \BS(T,0,X_0,I(T,k^*)) (\partial_{X_0}I(T,k^*))^2.
\]
Observe that
\[
\partial_{\sigma}\BS(T,0,X_0, I(T,k^*)) \partial_{X_0 X_0}^2I(T,k^*) =\sum_{j=1}^7 \tilde{I}_j(T).
\]The idea of the derivation of the curvature result is the following: In \cite{AlosLeon2017}, the authors derive the expression of the curvature for a model like \eqref{Black model} without jumps. Therefore, we will borrow their main result to deduce the contribution of the term $\tilde{I}_1(T)$. The contribution of the jumps will come from $\tilde{I}_2(T)$ and $\tilde{I}_4(T)$. The rest of the section is devoted to show that the asymptotics of $\tilde{I}_j(T)$ for $j = 3,5,6,7$ are negligible.
\subsection{Analysis of the integrals}
Regarding the first term, we borrow the result from \cite{AlosLeon2017}.
\begin{prop}
    Assume Hypotheses \ref{hyp 1} -- \ref{hyp 9} are met. Then we have
    \begin{align*}
        \lim_{T \to 0} \frac{\tilde{I}_1(T)}{T^{2H-\frac{1}{2}}} = &\frac{e^{X_0}}{\sqrt{2\pi}}\frac{\Sigma^2}{\sigma_0^3(H+\frac{1}{2})^2(2H+2)} \\
        + &\frac{\rho^2\Sigma^2}{\sigma_0^3(H+\frac{1}{2})^2}\left[ -\frac{6}{(H+\frac{3}{2})^2} + 2 B\left( H+\frac{3}{2}, H + \frac{3}{2}\right) + \frac{1}{2H+2}\right] \\
        + & \frac{2\rho^2 \Theta}{\sigma_0^2(H+\frac{1}{2})^2}B\left( H+\frac{3}{2}, H + \frac{3}{2}\right).
    \end{align*}
\end{prop}
\begin{proof}
    In \cite{AlosLeon2017} it is shown that, if the following limits
    \[
    \lim_{T \to 0} \frac{1}{T^{2H+2}}\E\left[\int_0^T \left( \E\left\{\int_r^T D_r \sigma_u^2 du | \mathcal{F}_r \right\}\right)^2 dr\right],
    \]
    \[
    \lim_{T \to 0} \frac{1}{T^{2+2H}} \E\left[  \int_0^T \int_s^T D_s\left\{ \sigma_r\int_s^T D_s\sigma_u^2 du \right\}drds\right],
    \]
    and
    \[
    \lim_{T \to 0} \frac{1}{T^{3+2H}} \E\left[ \left(\int_0^T\int_r^T D_r \sigma^2_u du dr \right)^2\right]
    \]
    are finite, then the ATM-IV curvature satisfies the following short-term asymptotic behaviour:
    \begin{align*}
        \lim_{T \to 0} \frac{\partial_{kk}^2I(T,k^*)}{T^{2H-1}} = &\frac{1}{4\sigma_t^2 }  \lim_{T \to 0} \frac{1}{T^{2H+2}}\E\left[\int_0^T \left( \E\left\{\int_r^T D_r \sigma_u^2 du | \mathcal{F}_r \right\}\right)^2 dr\right] \\
        -& \frac{3\rho^2}{2\sigma_t^5}\lim_{T \to 0} \frac{1}{T^{3+2H}} \E\left[ \left(\int_0^T\int_r^T D_r \sigma^2_u du dr \right)^2\right] \\
        + & \frac{\rho^2}{\sigma_t^4}\lim_{T \to 0} \frac{1}{T^{2+2H}} \E\left[  \int_0^T \int_s^T D_s\left\{ \sigma_r\int_s^T D_s\sigma_u^2 du \right\}drds\right].
    \end{align*}
    The result follows because Hypotheses \ref{hyp 6} and \ref{hyp 8} ensure the convergence of the limits and therefore, using that $D_s\sigma_t \sim \Sigma(t-s)^{H-\frac{1}{2}}$ and $D_rD_s\sigma_t = \Theta(t-s)^{H-\frac{1}{2}}(s-r)^{H-\frac{1}{2}}$, we can compute all the integrals to obtain the desired result.
\end{proof}
We now focus on the analysis of $\tilde{I}_2(T)$, whose contribution to the curvature is encapsulated in the following result.
\begin{prop}
    Assume Hypotheses \ref{hyp 1} -- \ref{hyp 9} are met. Then we have
    \[
    \lim_{T \to 0} \frac{\tilde{I}_2(T)}{T^{\frac{1-Y}{2}}} = \lim_{T \to 0} \frac{G(T,0,X_0, I^{(c)}(T,k^*)) - G(T,0,X_0, I(T,k^*))}{T^{\frac{1-Y}{2}}} = \frac{e^{X_0}}{\sqrt{2\pi}} \frac{(C_+ + C_{-})\Gamma(1-\frac{Y}{2})}{\sigma_0^{Y+1}Y(Y-1)2^{Y/2}}.
    \]
\end{prop}
\begin{proof}
    As $G$ is a continuously differentiable variable in the volatility argument, by the mean value theorem we can reduce the problem to the analysis of the difference between implied volatilities. To this end, consider $X^{(c)}$ a process satisfying model \eqref{Black model} without jumps. We write, for notation streamlining, $U_T = X_T^{(c)} - X_0$, $\hat{L}_t = L_t - \eta t$ and $g(x) = (e^x - 1)_+$. We also define the function
    \[
    F(x) = \E[g(U_T + x)].
    \]
    Notice that the ATM-call price under the continuous model $\mathcal{C}^{(c)}$ coincides with $e^{X_0}F(0)$ and the call price under the model with jumps $\mathcal{C}$ coincides with $e^{X_0}F(\hat{L}_T)$. Thus, the Itô formula for Lévy processes leads to
    \begin{equation} \label{eq: Levy price}
          \mathcal{C}- \mathcal{C}^{(c)} = \E\left[ \int_0^T \int_{\R} F(\hat{L}_{s-}+y) -F(\hat{L}_{s-}) -(e^y-1)F'(\hat{L}_{s-}) \nu(dy) ds\right].
    \end{equation}
    The strategy of the proof is the following:
    \begin{enumerate}
        \item Use the representation of Equation \eqref{eq: Levy price} to show that
        \[
        \lim_{T \to 0} \frac{\mathcal{C}-\mathcal{C}^{(c)}}{T^{\frac{3-Y}{2}}}
        \]
        is finite and determine its value.
        \item Show, using item 1, that the limit
        \[
        \lim_{T \to 0} \frac{I(T,k^*) - I^{(c)}(T,k^*)}{T^{1-\frac{Y}{2}}}
        \]
        is finite and determine its value.
        \item Using item 2 and the mean value theorem, find the limit
        \[
        \lim_{T \to 0} \frac{G(T,0,X_0,I(T,k^*)) - G(T,0,X_0, I^{(c)}(T,k^*))}{T^{\frac{1-Y}{2}}}
        \]
    \end{enumerate}
    Let's show first item 1. As in the proof of Proposition \ref{prop: I2}, the integral term in Equation \eqref{eq: Levy price} has to be split in the contribution of jumps of size $|y| \leq \delta$ and jumps of size $|y| > \delta$. In the same way, one can show that the term with jumps is of order $O(T)$. Hence, dividing by $T^{\frac{3-Y}{2}}$ shows that this term is of order $o(1)$, thus negligible. We are interested in the limit of 
    \[
   \E\left[ \int_0^T \int_{\R} F(\hat{L}_{s-}+y) -F(\hat{L}_{s-}) -(e^y-1)F'(\hat{L}_{s-}) \1_{\{|y| \leq \delta\}}\nu(dy) ds\right].
    \]
   as $T \to 0$, where the $o(1)$ term is the term involving the jumps of size $|y| \geq \delta$. Let $a_T = \sigma_0\sqrt{T}$, $s = \lambda T$ and define
    \[
    \hat{A}_{\lambda,T} = \frac{\hat{L}_{\lambda T}}{a_T}, \quad H(x,y) =\frac{1}{a_T}\left( F(a_T(x+y)) - F(a_Tx) - (e^{a_T y}-1)F'(a_T x)\right).
    \]
    Then the change of variables $s = \lambda T$, $y = a_T z$ and Fubini's theorem allows us to write 
    \[
    \frac{\mathcal{C}-\mathcal{C}^{(c)}}{e^{X_0}a_T^{1-Y}T}  =\int_0^1 \int_{\R} \E\left[H(R_{\lambda, T}, z) C\left(\frac{z}{|z|}\right)q(a_Tz)\1_{\{a_T|z| \leq \delta\}}|z|^{-1-Y}\right]dz d\lambda + o(1).
    \]
    where the $o(1)$ term is the one derived from the analysis of the term with jumps of size $|y| \geq \delta$. Let $\psi(x) = \phi(x)+ x\Phi(x)$ and let $\Psi(x) = \psi(x) - \psi(0) - \frac{x}{2}$. The following limits hold:
    \[
    \lim_{T \to 0} \frac{F(a_T x)}{a_T} = \psi(x), \quad \lim_{T \to 0}F'(a_T x) = \Phi(x).
.    \]
    Therefore,
    \[
    \lim_{T \to 0} H(R_{\lambda,T},z) =\Psi(z)
    \]
    in probability. The dominated convergence theorem and the fact that $q(x)$ is bounded as $x \to 0$ allows us to write
    \[
    \lim_{T \to 0} \frac{\mathcal{C}-\mathcal{C}^{(c)}}{T^{\frac{3-Y}{2}}} = e^{X_0}\sigma_0^{1-Y}\int_{\R} \Psi(z)C\left(\frac{z}{|z|}\right)|z|^{-1-Y}dz.
    \]
    Integrating by parts in the same way as with the ATM-IV skew shows that
    \[
    \lim_{T \to 0}\frac{\mathcal{C}-\mathcal{C}^{(c)}}{T^{\frac{3-Y}{2}}} =\frac{e^{X_0}}{\sqrt{2\pi}} \frac{(C_+ + C_{-})\sigma_0^{1-Y}\Gamma\left( 1-\frac{Y}{2}\right)}{Y(Y-1)2^{Y/2}},
    \]
    which proves Item 1. For Item 2, the mean value theorem and the fact that both $I(T,k^*)$ and $I^{(c)}(k^*)$ tend to $\sigma_0$ as $T \to 0$ allows us to write
    \begin{equation} \label{eq: item 2}
    \lim_{T \to 0}\frac{I(T,k^*) - I^{(c)}(T,k^*)}{T^{1-\frac{Y}{2}}}=\frac{(C_+ + C_{-})\sigma_0^{1-Y}\Gamma\left( 1-\frac{Y}{2}\right)}{Y(Y-1)2^{Y/2}}.
    \end{equation}
    Finally, for Item 3, the mean value theorem shows that there exists an intermediate point $\xi$ between $I(T,k^*)$ and $I^{(c)}(k^*)$ such that
    \[
    G(T,0,X_0, I(T,k^*)) - G(T,0,X_0,I^{(c)}(T,k^*)) = \partial_{\sigma}G(T,0,X_0, \xi)(I(T,k^*) - I^{(c)}(T,k^*)). 
    \]
    The proof is finished by combining the limit in Equation \eqref{eq: item 2} together with the fact that
    \[
    -\partial_{\sigma}G(T,0,X_0, \xi) \sim\frac{e^{X_0}}{\sigma_0^2\sqrt{2\pi}\sqrt{T}}
    \]
    as $T \to 0$, concluding therefore that
    \[
    \lim_{T \to 0} \frac{G(T,0,X_0, I^{(c)}(k^*)) - G(T,0,X_0, I(T,k^*))}{T^{\frac{1-Y}{2}}} = \frac{e^{X_0}}{\sqrt{2\pi}} \frac{(C_+ + C_{-})\Gamma(1-\frac{Y}{2})}{\sigma_0^{Y+1}Y(Y-1)2^{Y/2}}.
    \]
\end{proof}
The last contribution to the curvature is given by an analysis of the term $\tilde{I}_4(T)$.
\begin{prop}
     Assume Hypotheses \ref{hyp 1} -- \ref{hyp 9} are met. Then we have
    \[
    \lim_{T \to 0} \frac{\tilde{I}_4(T)}{T^{\frac{1-Y}{2}}} = -\frac{e^{X_0}}{\sqrt{2\pi}} \frac{(C_+ + C_ {-}) \Gamma\left(1-\frac{Y}{2} \right)}{\sigma_0^{Y+1}Y2^{Y/2}}.
    \]
\end{prop}
\begin{proof}
    Since $G = (\partial^2_{xx}-\partial_x)\BS$ and $\mathcal{B} = (\partial_x - \frac{1}{2})\BS$, we have the identity
    \[
    \tilde{I}_4(T) = \E\left[ \int_0^T \int_{\R} \Delta_{yy} G(T,s,X_s,v_s)\nu(dy) ds\right] + \E\left[\int_0^T \int_{\R} \Delta_{yy} \B(s,X_S,v_s) \nu(dy) ds \right].
    \]
    The second term has already been analysed in Proposition \ref{prop: I2} and therefore
    \[
    \E\left[\int_0^T \int_{\R} \Delta_{yy} \B(s,X_S,v_s) \nu(dy) ds \right] = O(T^{1-\frac{Y}{2}}).
    \]
    The conclusion of the result will arise from an asymptotic analysis on the term
    \[
    \tilde{I}^{(2)}_4(T) = \E\left[ \int_0^T \int_{\R} \Delta_{yy} G(T,s,X_s,v_s)\nu(dy) ds\right] .
    \]
    Again, we split the latter integral as
    \begin{align*}
    \tilde{I}^{(2)}_4(T) = &\E\left[ \int_0^T \int_{|y|\leq \delta}\Delta_{yy}G(T,s,X_s,v_s)\nu(dy) ds\right] + \E\left[ \int_0^T \int_{|y|> \delta}\Delta_{yy}G(T,s,X_s,v_s)\nu(dy) ds\right] \\
    =& \tilde{I}_4^{(2)}(T;\delta) +  \tilde{I}_4^{(2),c}(T;\delta).
    \end{align*}
    Recall that, under the definition $u_s = v_s\sqrt{T-s}$, we have
    \[
     G(T,s,X_s,v_s) = \frac{e^{X_0}}{u_s} \phi \left(\frac{X_s-X_0}{u_s} -\frac{u_s}{2}\right).
    \]
    Let $\mathcal{G}_t := \mathcal{F}^W_T \vee \mathcal{F}^L_T \vee \mathcal{F}_t$. Define also $\overline{G}_s(z) =\E\left[G(T,s,X_s+z,v_s) |\mathcal{G}_0 \right]$. Then, Lemma \ref{classic bound greeks} shows that
    \[
    \sup_{z \in \R} |\overline{G}_s(z)| \leq C T^{-1/2}, \quad \sup_{z \in \R} |\overline{G}_s'(z)| \leq C T^{-1}, \quad \sup_{z \in \R} |\overline{G}_s''(z)| \leq C T^{-3/2}.
    \]
    Hence
    \[
    \E[\Delta_{yy} G(T,s,X_s,v_s) | \mathcal{G}_0] = \overline{G}_s(y) - \overline{G}_s(0) - y \overline{G}_s'(0) = \Delta_{yy} \overline{G}_s(0).
    \]
    For jumps of size $|y| > \delta$ we have
    \begin{align*}
        |\tilde{I}_4^{(2),c}| \leq \int_{0}^T \left(2 \sup_{z \in \R} |\overline{G}_s(z)| \nu(\{|y| > \delta\}) + \sup_{z \in \R} |\overline{G}_s'(z)|\int_{|y| > \delta} |y|\nu(dy)\right)
    \end{align*}
    which implies that
    \[
    \lim_{T \to 0} \frac{\tilde{I}^{(2),c}_4(T)}{T^{\frac{1-Y}{2}}} = 0.
    \]
    We now deal with the small jumps contribution. Let $H(x,y)$ be defined as
    \[
    H(x,y ) = \phi(x+y) - \phi(x) - y \phi'(x).
    \]
    With the change of variables $s = \lambda T$, $y = u_{\lambda T} z$ and recalling that $A_{\lambda, T} = \frac{u_{\lambda T}}{\sqrt{T}}$ we have that
    \[
    \frac{\tilde{I}^{(2)}_4(T;\delta)}{T^{\frac{1-Y}{2}}} = e^{X_0}\int_0^1 \int_{\R}\E\left[  A_ {\lambda,T}^{-Y-1} H\left(\frac{X_{\lambda T}-X_0}{u_ {\lambda T}} - \frac{u_{\lambda T}}{2}, z \right)C\left(\frac{z}{|z|} \right)\1_{\{|u_{\lambda T}z| \leq \delta}|z|^{-1-Y} \right]dz d\lambda.
    \]
    Then,
    \[
    \left(A_{\lambda,T}, \frac{X_{\lambda T}-X_0}{u_ {\lambda T}} - \frac{u_{\lambda T}}{2}\right) \to (\sigma_0 \sqrt{1-\lambda}, Z_{\lambda})
    \]
    in distribution as $T \to 0$, where $Z_{\lambda} \sim N(0, \frac{\lambda}{1-\lambda})$. Since $q(u_{\lambda T}z) \to 1$ and $\1_{\{|u_{\lambda T}z|\leq \delta} \to 1$ as $T \to 0$ the dominated convergence theorem shows that
    \[
    \lim_{T \to 0} \frac{\tilde{I}^{(2)}_4(T;\delta)}{T^{\frac{1-Y}{2}}}  = \int_0^1 \int_{\R}\frac{e^{X_0}}{\sigma_0^{Y+1}(1-\lambda)^{\frac{Y+1}{2}}}\E[H(Z_{\lambda}, z)] C\left( \frac{z}{|z|}\right)|z|^{-1-Y} dz d\lambda.
    \]
    In the same way as for the ATM-IV skew, we have
    \[
    \E\left[ H(Z_{\lambda},z)\right] = \sqrt{1-\lambda}\left(\phi(z\sqrt{1-\lambda}) - \phi(0) \right).
    \]
    The change of variables $w = z\sqrt{1-\lambda}$ gives
    \[
    \int_{\R} \E[H(Z_{\lambda},z)] C\left( \frac{z}{|z|}\right) |z|^{-1-Y} dz  = (C_+ + C_{-})(1-\lambda)^{\frac{Y+1}{2}}\int_0^{\infty}(\phi(w) - \phi(0))w^{-1-Y} dw.
    \]
    Thus,
    \[
    \lim_{T \to 0}\frac{\tilde{I}^{(2)}_4(T;\delta)}{T^{\frac{1-Y}{2}}} = \frac{e^{X_0}(C_+ + C_{-})}{\sigma_0^{Y+1}} \int_{\R} (\phi(w) - \phi(0)) w^{-1-Y}dw.
    \]
    Integrating by parts we end up with
    \[
     \lim_{T \to 0}\frac{\tilde{I}^{(2)}_4(T;\delta)}{T^{\frac{1-Y}{2}}} = -\frac{e^{X_0}(C_+ + C_{-})\Gamma\left( 1-\frac{Y}{2}\right)}{\sqrt{2\pi}\sigma_0^{Y+1}Y 2^{Y/2}},
    \]
    concluding the asymptotic behaviour of $\tilde{I}_4(T)$.
\end{proof}

The remaining of the section is devoted to show that the other terms are negligible. We start with $\tilde{I}_3(T)$.
\begin{prop}
     Assume Hypotheses \ref{hyp 1} -- \ref{hyp 9} are met. Then we have
    \[
    \tilde{I}_3(T) = \frac{\rho}{2}\E\left[\int_0^T (\partial^2_{xx}-\frac{1}{2})\partial_x(G(T,s,X_s,v_s)  - G(T,s,X^{(c)}_s,v_s))\Lambda_s ds \right] = o\left(T^{\min\left\{ 2H-\frac{1}{2}, \frac{1-Y}{2}\right\}} \right).
    \]
\end{prop}
\begin{proof}
    Let $\hat{L}_t = L_t - \eta t$ and let $F = (\partial^2_{xx} - \frac{1}{2})\partial_xG$. Then, we write $\tilde{I}_3(T)$ as
    \[
    \tilde{I}_3(T) = \frac{\rho}{2} \int_0^T \E[\Lambda_s(F(s,X_s^{(c)}+\hat{L}_s, v_s) - F(s,X_s^{(c)}, v_s)] ds.
    \]
   Let
\[
A_s :=
F(s,X_s^{(c)}+\hat L_s,v_s)
-
F(s,X_s^{(c)},v_s).
\]
For every fixed \(z\in\mathbb R\), define
\[
\overline F_s(z)
:=
\E\left[
F(s,X_s^{(c)}+z,v_s)
\mid\mathcal F_T^W
\right].
\]
Conditioning first on
\(\mathcal F_T^W\vee\mathcal F_T^L\), we obtain
\[
\E\left[
A_s
\mid
\mathcal F_T^W\vee\mathcal F_T^L
\right]
=
\overline F_s(\hat L_s)-\overline F_s(0).
\]
Since \(\Lambda_s\) is \(\mathcal F_T^W\)-measurable, the tower
property gives
\[
\E[\Lambda_s A_s]
=
\E\left[
\Lambda_s
\left\{
\overline F_s(\hat L_s)-\overline F_s(0)
\right\}
\right].
\]
Conditionally on \(\mathcal F_T^W\), the function
\(z\mapsto\overline F_s(z)\) is deterministic. Hence Dynkin's
formula yields
\[
\E\left[
\overline F_s(\hat L_s)-\overline F_s(0)
\mid\mathcal F_T^W
\right]
=
\int_0^s
\E\left[
(\hat{\mathcal L}\overline F_s)(\hat L_r)
\mid\mathcal F_T^W
\right]dr.
\]
    The bound of the difference $ \overline{F}_s(\hat{L}_s) - \overline{F}_s(0) $ is therefore a problem of bounding the first two derivatives of the function $\overline{F}_s$. Towards this end we decompose $X^{(c)}$ as
    \[
    X^{(c)}_t = \xi^W_t + \xi^B_t,
    \]
    where
    \[
    \xi_t^W = X_0 - \frac{1}{2}\int_0^t \sigma_s^2 + \rho\int_0^t \sigma_s dW_s \in \mathcal{F}^W_T, \quad \xi^B_t = \sqrt{1-\rho^2}\int_0^t\sigma_s dB_s.
    \]
    Conditionally on $\mathcal{F}_T^W$, $\xi^B_s | \mathcal{F}^W_T \sim N(0, \varsigma_s^2)$, where
    \[
    \varsigma_s^2 = (1-\rho^2) \int_0^s \sigma_s^2 ds.
    \]
    Observe that $G(T,s,X_s, v_s) = e^{X_0} p_{u_s^2}\left(X_s-X_0 - \frac{u_s^2}{2}\right)$ where $p_a(x)$ denotes the density of a $N(0,a)$ random variable evaluated at $x$. Then, if we define $\overline{G}_s(z) = \E[G(T,s,X^{(c)}+z,v_s)| \mathcal{F}^W_T]$, we have that
    \begin{align*}
    \overline{G}_s(z) = &\E\left[G(T,s,\xi^W_s + \xi^B_s + z, v_s) | \mathcal{F}^W_T \right] \\
    = & e^{X_0}\int_{\R} p_ {u_s^2}(\xi_s^W + z-X_0-\frac{u_s^2}{2} + \theta)p_{\varsigma_s^2}(\theta) d\theta \\
    = & e^{X_0}p_{u_s^2 + \varsigma_s^2}(\xi_s^W + z - X_0 - \frac{u_s^2}{2}).
    \end{align*}
    Let $w^2_s = u_s^2 + \varsigma_s^2$.
    This identity for $\overline{G}_s$ allows us to write
    \[
    \overline{F}_s(z) = e^{X_0}\left(p^{(3)}_{w_s^2}\left(\xi_s^W - X_0 - \frac{u_s^2}{2} + z\right) -\frac{1}{2}p_{w_s^2}'\left( \xi_s^W - X_0 - \frac{u_s^2}{2} + z\right) \right).
    \]
    Using classical bounds for the density function of the standard normal distribution we are able to write
    \[
    \sup_{z \in \R} |\overline{F}_s(z)| \leq C\left(\frac{1}{w_s^4} + \frac{1}{w_s^2} \right), \quad \sup_{z \in \R} |\overline{F}_s'(z) | \leq C\left(\frac{1}{w_s^5} + \frac{1}{w_s^3} \right),
    \]
    and
    \[
    \sup_{z \in \R} |\overline{F}''_s(z)| \leq C\left( \frac{1}{w_s^{6}} + \frac{1}{w_s^4}\right).
    \]
    Since the volatility process $\sigma$ is bounded from above and below, there exists constants $C_1, C_2 > 0$ with $C_1 < C_2$ such that
    \[
    C_1 \sqrt{T} \leq w_s \leq C_2 \sqrt{T}.
    \]
    Hence,
    \[
     \sup_{z \in \R} |\overline{F}_s(z)| \leq O(T^{-2}), \quad \sup_{z \in \R} |\overline{F}_s'(z) |\leq O(T^{-\frac{5}{2}}), \quad  \sup_{z \in \R} |\overline{F}''_s(z)| \leq O(T^{-3}).
    \]
    This bounds, together with Hypothesis \ref{hyp 1} allows us to obtain a bound for $\hat{\mathcal{L}}\overline{F}_s(z)$. To do so, we will split the integral into three regions, obtain a bound for each region and deduce the global bound. In order to construct the three regions, let $\delta \in (0,1)$ an such that, without loss of generality, $\sqrt{T} \leq \delta$ and $\nu(dy) \leq C|y|^{-1-Y}$ for $0<|y|\leq \delta$. On the region $|y| \leq \sqrt{T}$ we have that $|e^y-1-y| \leq C y^2$, so the bound on the generator in that region can be computed as
    \begin{align*}
    &\left|\int_{|y| \leq \sqrt{T}}(\overline{F}_s(z+y) - \overline{F}_s(z) - y \overline{F}_s'(z)) \nu(dy) -\int_{|y|\leq \sqrt{T}} (e^y-1-y) \overline{F}'_s(z) \nu(dy)\right| \\
    \leq & C\left(\sup_{z \in \R} |\overline{F}''_s(z)| + \sup_{z \in \R} |\overline{F}'_s(z)| \right)\int_{|y|\leq \sqrt{T}} y^2 \nu(dy) \\
    \leq & C T^{-2-\frac{Y}{2}}.
    \end{align*}
    On the region $\sqrt{T}< |y| \leq \delta$ the bound is more elementary, since
    \begin{align*}
         &\left|\int_{\sqrt{T} <|y| \leq \delta}(\overline{F}_s(z+y) - \overline{F}_s(z) - y \overline{F}_s'(z)) \nu(dy) -\int_{\sqrt{T} <|y| \leq \delta} (e^y-1-y) \overline{F}'_s(z) \nu(dy)\right| \\
         \leq &2\sup_{z \in \R} |\overline{F}_s(z)|\int_{\sqrt{T}<|y| \leq \delta}\nu(dy) +  \sup_{z \in \R} |\overline{F}'_s(z)| \int_{\sqrt{T}< |y| \leq \delta} |e^y - 1| \nu(dy) \\
         \leq & C T^{-2- \frac{Y}{2}},
    \end{align*}
    where we have used that there exists $C > 0$ such that $|e^y - 1| \leq C|y|$ for $\sqrt{T} < |y| \leq \delta$. Finally, in the region $|y| > \delta$ we use the same bounding strategy as before, that is,
    \begin{align*}
     &\left|\int_{|y| > \delta}(\overline{F}_s(z+y) - \overline{F}_s(z) - y \overline{F}_s'(z)) \nu(dy) -\int_{|y| > \delta} (e^y-1-y) \overline{F}'_s(z) \nu(dy)\right| \\
      \leq &2\sup_{z \in \R} |\overline{F}_s(z)|\int_{|y| > \delta}\nu(dy) +  \sup_{z \in \R} |\overline{F}'_s(z)| \int_{|y| > \delta} |e^y - 1| \nu(dy) \\
      \leq & CT^{-2-\frac{Y}{2}}.
    \end{align*}
    In this case, we have used that $Y > 1$ and $T < 1$ to obtain the desired power. Back to Dynkin's formula, the previous bounds allow us to derive the estimate
      \begin{align*}
     \E\left[\overline{F}_s(\hat{L}_s) - \overline{F}_s(0) | \mathcal{F}^W_T\right] =  &\int_0^s \E\left[\hat{\mathcal{L}}\overline{F}_s(\hat{L}_r) | \mathcal{F}^W_T \right] dr \\
     \leq & s \sup_{z \in \R} \E[\hat{\mathcal{L}}\overline{F_s}(z) | \mathcal{F}^W_T] \\
      \leq& C s T^{-2-\frac{Y}{2}},
    \end{align*}
    where $C > 0$ denotes a constant that does not depend on $T$. Thus, using the hypotheses on the Malliavin derivative of $\sigma$, we have
    \[
    \E\left[ |\Lambda_s| \cdot\left| \E\left[\overline{F}_s(\hat{L}_s) - \overline{F}_s(0) | \mathcal{F}^W_T \right] \right| \right] \leq C s(T-s)^{H+\frac{1}{2}}T^{-2-\frac{Y}{2}}.
    \]
    Finally,
    \begin{align*}
         |\tilde{I}_3(T)|\leq & C \int_0^T s(T-s)^{H+\frac{1}{2}}T^{-2-\frac{Y}{2}} ds \\
         \leq & C T^{H + \frac{1}{2}- \frac{Y}{2}}.
    \end{align*}
    Since $H > 0$, we have
    \[
    H + \frac{1-Y}{2} > \frac{1-Y}{2} \geq \min \left\{ 2H-\frac{1}{2}, \frac{1-Y}{2}\right\},
    \]
    so
    \[
     |\tilde{I}_3(T)| \leq o\left(T^{\min\left\{ 2H-\frac{1}{2}, \frac{1-Y}{2}\right\}} \right)
    \]
    concluding therefore the result.
\end{proof}
The next results are straightforward, since they rely on previous estimates and the result from Theorem \ref{th: main result} about the ATM-IV skew.
\begin{prop}
     Assume Hypotheses \ref{hyp 1} -- \ref{hyp 9} are met. Then we have
    \[
    \tilde{I}_5(T) = -c_2 \E\left[ \int_0^T (\partial^2_{xx}-\frac{1}{2}) \partial_x \BS(T,s,X_s,v_s) ds\right] =  o\left(T^{\min\left\{ 2H-\frac{1}{2}, \frac{1-Y}{2}\right\}} \right).
    \]
\end{prop}
\begin{proof}
        We have
        \[
        \left(\partial^2_{xx} - \frac{1}{2} \right)\partial_x \BS  = \partial_x G + G + \frac{1}{2}\partial_x \BS.
        \]
        Using Lemma \ref{classic bound greeks} and the fact that
        \[
        0 \leq \partial_x \BS(T,s.X_s,v_s) \leq e^{X_s}, \quad \E[e^{X_s}] = e^{X_0}
        \]
        we find that there exists a constant $C > 0$ such that
        \[
        |\tilde{I}_5(T)| \leq C\int_0^T\left(\frac{1}{T} + \frac{1}{\sqrt{T}} + 1\right) dt =O(1),
        \]
        so 
        \[
        |\tilde{I}_5(T)| \leq o\left(T^{\min\left\{ 2H-\frac{1}{2}, \frac{1-Y}{2}\right\}} \right)
        \]
    \end{proof}
    For $\tilde{I}_6(T)$ and $\tilde{I}_7(T)$ we use the skew asymptotics in Theorem \ref{th: main result}.
    \begin{prop}
         Assume Hypotheses \ref{hyp 1} -- \ref{hyp 9} are met. Then we have
        \[
        \tilde{I}_6(T) = -2\partial_{x\sigma}\BS(T,0,X_0,I(T,k^*))\partial_{X_0}I(T,k^*)  =  o\left(T^{\min\left\{ 2H-\frac{1}{2}, \frac{1-Y}{2}\right\}} \right).
        \]
    \end{prop}
    \begin{proof}
        We use the at-the-money identity
        \[
        \partial_{x\sigma}\BS(T,t,x,\sigma) = \frac{1}{2}\partial_{\sigma} \BS(T,t,x,\sigma).
        \]
        Hence, using Theorem \ref{th: main result} we have that
        \[
        \tilde{I}_6(T) = O\left(T^{^{\min\{H, 1-\frac{Y}{2}\}}} \right) =o\left(T^{\min\left\{ 2H-\frac{1}{2}, \frac{1-Y}{2}\right\}} \right).
        \]
    \end{proof}
    \begin{prop}
         Assume Hypotheses \ref{hyp 1} -- \ref{hyp 9} are met. Then we have
        \[
        \tilde{I}_7(T) =  - \partial_{\sigma\sigma} \BS(T,0,X_0,I(T,k^*)) (\partial_{X_0}I(T,k^*))^2 = o\left(T^{\min\left\{ 2H-\frac{1}{2}, \frac{1-Y}{2}\right\}} \right).
        \]
    \end{prop}
    \begin{proof}
        We use the at-the-money identity
        \begin{align*}
        \partial_{\sigma\sigma} \BS(T,0,X_0,I(T,k^*)) (\partial_{X_0}I(T,k^*))^2 = &\frac{I(T,k^*) T}{4}\partial_{\sigma}\BS(T,0,X_0,I(T,k^*)) (\partial_{X_0}I(T,k^*))^2\\
        = &O\left(T^{\min\{2H+\frac{3}{2}, \frac{5}{2}-Y\}} \right).
        \end{align*}
        Thus,
        \[
        \lim_{T \to 0} \tilde{I}_7(T) = 0
        \]
        so, in particular, the conclusion of the proposition holds. 
    \end{proof}
   These are all the necessary ingredients to prove Theorem \ref{th: main result 2}.
    \begin{proof}(Theorem \ref{th: main result 2})
    Assume first that
    \[
    2H - 1 < -\frac{Y}{2}.
    \]
    Then, 
    \[
    \lim_{T \to 0} \frac{\partial_{\sigma} \BS(T,T,0,X_0, I(T,k^*)) \partial^2_{kk}I(T,k^*)}{T^{2H-\frac{1}{2}}} = \lim_{T \to 0} \frac{\tilde{I}_1(T)}{T^{2H-\frac{1}{2}}}=\frac{e^{X_0}}{\sqrt{2\pi}} C_{vol},
    \]
    for $C_{vol}$ defined in Theorem \ref{th: main result 2}. Then, since
    \[
    \frac{\partial_{\sigma} \BS(T,T,0,X_0, I(T,k^*))}{\sqrt{T}} \sim \frac{e^{X_0}}{\sqrt{2\pi}}
    \]
    we have that
    \begin{align*}
         \lim_{T \to 0} \frac{\partial_{\sigma} \BS(T,0,X_0, I(T,k^*)) \partial^2_{kk}I(T,k^*)}{T^{2H-\frac{1}{2}}} =   \lim_{T \to 0} \frac{e^{X_0} \partial^2_{kk}I(T,k^*)}{\sqrt{2\pi}T^{2H-1}} = \frac{e^{X_0}}{\sqrt{2\pi}}C_{vol},
    \end{align*}
    which concludes the result in the case $2H-1 < -\frac{Y}{2}$. The case $2H-1 > - \frac{Y}{2}$ follows the same lines, taking into account that, in this case,
    \[
    \lim_{T \to 0} \frac{\partial_{\sigma}\BS(T,0,X_0, I(T,k^*))\partial^2_{kk}I(T,k^*)}{T^{\frac{1-Y}{2}}} = \lim_{T \to 0} \frac{\tilde{I}_2(T) + \tilde{I}_4(T)}{T^{\frac{1-Y}{2}}} = \frac{e^{X_0}}{\sqrt{2\pi}} C_{jumps}
    \]
    for $C_{jumps}$ defined in Theorem \ref{th: main result 2}. The case $2H-1 > - \frac{Y}{2}$ then follows from cancelling the $\frac{e^{X_0}}{\sqrt{2\pi}}$ from the short-term limit of $\partial_{\sigma}\BS(T,0,X_0,k,I(T,k))$.

    The critical exponent case $2H-1 = -\frac{Y}{2}$ follows trivially from the other two cases.
    \end{proof}
\section{An example: The Rough Heston model} \label{sec 5}
The rough Heston model with stable jumps assumes that the log-price $X$ is of the form
\[
X_t = X_0 - \eta t - \frac{1}{2}\int_0^t V_s ds + \int_0^t \sqrt{V_s}(\rho dW_s + \sqrt{1-\rho^2}dB_s) + L_t,
\]
where
\[
V_t = V_0 + \kappa \int_0^t K(t-s) (\theta-V_s) ds + \nu\int_0^t K(t-s) \sqrt{V_s} dW_s, \quad K(t) = \frac{1}{\Gamma\left(H+\frac{1}{2}    \right)}t^{{H-\frac{1}{2}}}\1_{\{t > 0\}}.
\]
Instead of showing that $V$ fits into Hypotheses \ref{hyp 2} -- \ref{hyp 6} we will find an approximation of $V$ satisfying the hypothesis and with the same short-time behaviour as $V$. Consider its approximation
\[
V_t^n = V_0 + \kappa \int_0^t K(t-s) (\theta-b_n(V^n_s)) ds + \nu\int_0^t K(t-s)\sqrt{\sigma_n(V^n_s)}dW_s
\]
where $b_n$ is a monotonically increasing smooth function satisfying $b_n(x) = x$ if $x \in [0, n]$ and $b_n(x) \to n^2$ if $x \to \infty$, and $\sigma_n$ is a monotonically increasing smooth function such that $\sigma_n(x) = x$ if $n^{-1} \leq x \leq n$ and $n^{-2} \leq \sigma_n(x) \leq n^2$ for all $x \in [0, \infty)$. Notice that, if we define the stopping times
\[
\tau_n = \inf\{ t > 0; V_t \notin[1/n, n]\},
\]
then
\[
\{V_t \neq V_t^n \}\subseteq \{\tau_n \leq T\}.
\]
Thus, it is convenient to analyse the stopping time $\tau_n$ and the stopped variance process $V_{t \wedge \tau_n}$ in order to perform an error analysis between $V$ and $V^n$. The first thing we shall prove regarding the regularity of $V$ is the following result
\begin{prop} \label{prop: Integrable variance}
    The variance process $V$ belongs to $L^1(\Omega)$ for every $t \in [0,T]$. Moreover, if $m(t) = \E[V_t]$, it holds that
    \[
    \int_0^T m(t) dt < \infty.
    \]
\end{prop}
\begin{proof}
    We borrow the Lemma 3.1 in \cite{AbiJaber2019} which shows that in particular, for the rough Heston case, one has
    \[
    \sup_{t \in [0,T]} \E[V_t] < \infty.
    \]
    The result therefore is deduced from this fact.
\end{proof}
Observe that a simple consequence of this fact is that the fair strike of the volatility swap under the rough Heston model is well defined. Indeed, using  Jensen's inequality and Fubini's theorem we show that
\[
\E[v_0] := \E\left[\sqrt{\frac{1}{T}\int_0^T  V_t dt} \right] \leq \sqrt{\frac{1}{T}\int_0^T \E[V_t] dt}, 
\]
which is finite from Proposition \ref{prop: Integrable variance}. 

In the remaining of this section, we will prove several inequalities that will illustrate how close the approximated process $V^n_t$ is from the original rough Heston variance process $V$. These results will allow us to justify that the short-time asymptotics of both processes coincide.
\begin{lema}
    Let $h$ be an adapted process with continuous sample paths uniformly bounded in $\omega \in \Omega$ by a constant $L > 0$. Let
    \[
    M_t = \int_0^t K(t-s)h_s dW_s.
    \]
    Then, for every $p \geq 2$ there exists a constant $C_p >0 $ depending on $p$ and $H$ such that
    \[
    \E\left[\sup_{t \in [0,T]} |M_t|^p\right] \leq C_p L^p T^{pH}.
    \]
\end{lema}
\begin{proof}
    From now on, $C_p >0$ will denote a constant that may change from line to line. Observe that
    \[
    M_t - M_s = \int_0^s (K(t-u) - K(s-u))h_u  dW_s + \int_s^t K(t-u)h_u dW_u.
    \]
    Therefore, by the Burkholder-Davis-Gundy inequality in \cite{Pannier2025} we have, for every $q \geq 2$,
    \[
    \E[|M_t - M_s|^q] \leq C_q L^q \left( \int_0^s (K(t-u) - K(s-u))^2 du + \int_s^t K(t-u)^2 du\right)^{q/2}.
    \]
    Using the definition of $K$ and a change of variables, we can show that there exists a constant $C_q > 0$  such that
    \[
     \E[|M_t - M_s|^q] \leq C_q L^q |t-s|^{qH}.
    \]
    Fix $p \geq 2$ and choose
    \[
    q > p \vee \frac{1}{H}.
    \]
    Choose also $\gamma$ such that 
    \[
    \frac{1}{p}< \gamma < H.
    \]
    Therefore, the Garsia-Rodemich-Rumsey lemma shows that
    \[
    \sup_{t \in [0,T]} |M_t|^q \leq C_qT^{\gamma q-1} \int_0^T \int_0^T \frac{|M_t - M_s|^q}{|t-s|^{1+\gamma q}} ds dt.
    \]
    Taking expectations and using the moment estimate for the moments of $|M_t|$ we have
   \[
   \E\left[ \sup_{t \in [0,T]} |M_t|^q\right] \leq C_q L^q T^{\gamma q - 1}\int_0^T \int_0^T |t-s|^{qH-1-\gamma q}dsdt.
   \]
   Thus,
   \[
     \E\left[ \sup_{t \in [0,T]} |M_t|^q\right] \leq C_q L^q T^{qH}.
   \]
   Finally, since $q \geq p$ we have
   \[
     \E\left[ \sup_{t \in [0,T]} |M_t|^p\right] \leq \left(   \E\left[ \sup_{t \in [0,T]} |M_t|^q\right]\right)^{p/q} \leq C_p L^p T^{pH},
   \]
    as desired.
\end{proof} 
The previous result helps us to control the error between $V^n$ and $V$ as it will allow us to deal with the stochastic part. The conclusion of the error bound is the following proposition.
\begin{prop} \label{prop: moment sup diference}
    For every $p \geq 2$ there exists a constant $C_{p,n} >0$ such that, for every $T \leq 1$,
    \[
    \E \left[ \sup_{t \in [0,T]} |V_t^n - V_0|^p\right] \leq C_{p,n} T^{pH}.
    \]
\end{prop}
\begin{proof}
    We can write the difference $V_t^n - V_0 = D_t + M_t$ where
    \[
    D_t = \kappa \int_0^t K(t-s)(\theta-b_n(V^n_s))ds, \quad M_t = \nu \int_0^t K(t-s)\sqrt{\sigma_n(V^n_s)}dW_s.
    \]
    Using that $|b_n(x)| \leq n^2$ we can easily find
    \[
    |D_t| \leq\kappa(\theta + n^2) \frac{T^{H+\frac{1}{2}}}{\Gamma(H+\frac{3}{2})}.
    \]
    This, together with the previous lemma, gives the conclusion.
\end{proof}
Using these results, we can analyse the stopping time law.
\begin{prop}
    Let $n$ be big enough so that $\delta_n = \min(V_0 - 1/n, n-V_0) > 0$. Then, there exists $C >0$ such that for all $p \geq 2$,
    \[
    \P(\tau_n \leq T) \leq C \delta_n^{-p} T^{pH}.
    \]
\end{prop}
\begin{proof}
    If $\tau_n \leq T$ then $\sup_{t \in [0,T]} |V^n_t - V_0| \geq \delta_n$, so
    \[
    \P(\tau_n \leq T) \leq \P\left(\sup_{t \in [0,T]} |V^n_t - V_0| \geq \delta_n \right) \leq \frac{\E\left[ \sup_{t \in [0,T]} |V_t^n - V_0|^p\right]}{\delta_n^p}.
    \]
    Thus, by the previous analysis, there exists $C > 0$ such that for all $p \geq 2$ we have
    \[
     \P(\tau_n \leq T) \leq C \delta_n^{-p} T^{pH},
    \]
    as desired.
\end{proof}
So far we have approximated $V$ by a more regular process $V^n$. Even though this process will satisfy the majority of hypotheses, it does not satisfy that there exist two constants $c_1, c_2 > 0$ such that $c_1 \leq V_t^n \leq c_2$. To solve this problem, we will assume that the volatility process of $X$ is $\sqrt{V^{n,m}_t}:=\sqrt{\sigma_m(V^n_t)}$ where $\sigma_m$ has been defined previously. Since $\sigma_m$ and all its derivatives are bounded, Hypothesis \ref{hyp 2} is satisfied and if Hypotheses \ref{hyp 3} -- \ref{hyp 9} are satisfied for $V^n$, then they are also satisfied for $V^{n,m}$. Therefore, an analysis of the Malliavin derivatives of $V^n$ will give us that $V^{n,m}$ satisfies Hypotheses \ref{hyp 2} -- \ref{hyp 9} and both Theorem \ref{th: main result} and Theorem \ref{th: main result 2} can be applied to this approximated setup.

The reason why we introduce this approximated rough Heston variance is because, thanks to Propositions \ref{prop: diff iv's}, \ref{prop: diff skews} and \ref{prop: diff curvatures} below, we can ensure that the short-time behaviour of the original rough Heston variance and our approximation is the same.

\begin{prop} \label{prop: diff iv's}
    Let $\mathcal{C}$ be the price of the call option under $V$ and let $\mathcal{C}^{n}$ be the price of the call option under $V^{n} := V^{n,n}$. Let $I^n$ and $I$ be the implied volatilities for the models with the approximated and the original variance respectively. Then, for every $m \in \N$ we have
    \[
    \lim_{T \to 0} \frac{|I(T,k^*) - I^n(T,k^*)|}{T^m} = 0.
    \]
\end{prop}
\begin{proof}
    Without loss of generality, replace $\tau_n$ by $\tau_n \wedge T$, so that $\tau_n$ is a bounded stopping time. The mean value theorem shows that there exists an intermediate value $\xi$ such that
    \[
    |I(T,k^*) - I^n(T,k^*)|= \frac{|\mathcal{C} - \mathcal{C}^n|}{\partial_{\sigma}\BS(T,0,X_0,\xi)}.
    \]
    Since we are interested in the short time behaviour of the previous expression and there exists a constant $c > 0$ such that
    \[
    \lim_{T \to 0} \frac{\partial_{\sigma}\BS(T,0,X_0,\xi)}{c\sqrt{T}} = 1,
    \]
    it is enough to study the quotient
    \[
    \frac{|\mathcal{C} - \mathcal{C}^n|}{\sqrt{T}}.
    \]
    On the one hand, since the map $x \mapsto (x-k)_+$ is Lipschitz continuous with Lipschitz constant one, we can write
    \[
    |\mathcal{C} - \mathcal{C}^n| \leq \E[|S_T-S_T^n|\1_{\{\tau_n \leq T\}}].
    \]
    The optimal sampling theorem allows us to write
    \[
    |\mathcal{C} - \mathcal{C}^n|\leq 2\E[|S_{\tau_n}| \1_{\{\tau_n \leq T\}}].
    \]
    Since the stopped variance is bounded, there exists $v(n)$ such that $V_{t \wedge \tau_n} \leq v(n)$ then
    \[
    \E[|S_{\tau_n}|^q] \leq S_0^q \exp\left( \frac{q(q-1)}{2}v(n) T + T \psi_L(q)\right),
    \]
    where $\psi_L(q)$ is the exponent defined as
    \[
    \E[e^{q(L_t-\eta t)}] = e^{\psi_L(q)t}.
    \]
    Consequently,
    \[
    \sup_{T \in (0,1]} \E[|S_{\tau_n}|^q]  < \infty.
    \]
    Since
    \[
    \P(\tau_n \leq T) \leq C_{n,p} T^{pH},
    \]
    we can apply Hölder's inequality to show that
    \[
    |\mathcal{C}- \mathcal{C}^n|\leq 2\sup_{T \in (0,1]} \E[|S_{\tau_n}|^q]^{1/q}\P(\tau_n \leq T)^{1-\frac{1}{q}} \leq C_{q,p,n} T^{p(1-\frac{1}{q})H}.
    \]
    As $p$ can be chosen arbitrarily large, we have that
    \[
    \frac{|\mathcal{C}- \mathcal{C}^n|}{\sqrt{T}} \leq C_{q,p,n}T^{p(1-\frac{1}{q})H-\frac{1}{2}} 
    \]
    and therefore, for a fixed $m \in \N$,
    \[
    0 \leq \frac{|I(T,k^*)-I^n(T,k^*)|}{T^m} \leq CT^{p(1-\frac{1}{q})H-\frac{1}{2}-m}
    \]
    which goes to zero as $T \to 0$ because, again, $p$ is chosen arbitrarily large.
\end{proof}
A similar result can be obtained for the ATM-IV skew.
\begin{prop} \label{prop: diff skews}
    Let $\mathcal{C}$ be the price of the call option under $V$ and let $\mathcal{C}^{n}$ be the price of the call option under $V^{n} := V^{n,n}$. Let $I^n$ and $I$ be the implied volatilities for the models with the approximated and the original variance respectively. Then, for every $m \in \N$ we have
    \[
    \lim_{T \to 0} \frac{|\partial_kI(T,k^*) - \partial_kI^n(T,k^*)|}{T^m} = 0.
    \]
\end{prop}
\begin{proof}
    Implicit differentiation with respect to the strike variable shows that
    \[
    \partial_k \mathcal{C}= \partial_k\BS(T,0,X_0,I(T,k^*)) + \partial_{\sigma} \BS(T,0,X_0, I(T,k^*)) \partial_kI(T,k^*).
    \]
    This implies that
    \[
    \partial_kI(T,k^*) = \frac{\Phi \left(-\frac{I(T,k^*)\sqrt{T}}{2} \right)- \P(X_T > X_0)}{\sqrt{T} \phi\left( \frac{I(T,k^*)\sqrt{T}}{2}\right)}.
    \]
    Let 
    \[
    \mathcal{S}(x,\sigma)=  \frac{\Phi \left(-\frac{\sigma\sqrt{T}}{2} \right)- x}{\sqrt{T} \phi\left( \frac{\sigma\sqrt{T}}{2}\right)}.
    \]
    We have
    \[
    \partial_kI(T,k^*) = \mathcal{S}(\P(X_T > X_0), I(T,k^*)), \quad \partial_kI^n(k^*) = \mathcal{S}(\P(X^n_T > X_0), I^n(T,k^*)),
    \]
    so, 
    \begin{align*}
         &\partial_kI(T,k^*)  - \partial_kI^n(T,k^*)\\
         =  &\mathcal{S}(\P(X_T > X_0), I(T,k^*)) - \mathcal{S}(\P(X^n_T > X_0), I^n(T,k^*)) \\
         = &\partial_x\mathcal{S}(\xi_1, I(T,k^*))(\P(X_T > X_0) - \P(X_T^n > X_0)) + \partial_{\sigma}\mathcal{S}(\P(X^n_T > X_0, \xi_2)(I(T,k^*) - I^n(T,k^*)).
    \end{align*}
    The first observation that we do is that, since $I(T,k^*), I^n(T,k^*) \to \sqrt{V_0}$ as $T \to 0$, both quantities lie on a compact set for $T$ small enough. Next, elementary computations show that
    \[
    \partial_x \mathcal{S}(x,\sigma) = \frac{-1}{\sqrt{T}\phi\left(\frac{\sigma\sqrt{T}}{2} \right)}, \quad \partial_{\sigma} \mathcal{S}(x,\sigma) = \frac{-1}{2}+ \frac{\sigma \sqrt{T}}{4\phi\left( \frac{\sigma \sqrt{T}}{2}\right)}\left( \Phi\left(\frac{-\sigma\sqrt{T}}{2}\right)-x \right),
    \]so by taking $T \to 0$ we see that the derivatives of $\mathcal{S}$ behave as
    \[
    \partial_x \mathcal{S} = O(\sqrt{T}), \quad \partial_{\sigma} \mathcal{S} = O(1),
    \]
    so the difference of skews as $T \to 0$ is comparable to
    \[
    |\partial_kI(T,k^*) - \partial_kI^n(T,k^*)| \sim |I(T,k^*) - I^n(T,k^*)| + \frac{|\P(X_T > X_0) - \P(X_T^n > X_0)|}{\sqrt{T}}.
    \]
    Regarding the second term, notice that
    \[
    \{\1_{\{X_T > X_0\}} \neq \1_{\{X_T^n > X_0\}} \} \subseteq \{\tau_n \leq T\},
    \]
    so 
    \[
    \frac{|\P(X_T > X_0) - \P(X_T^n > X_0)|}{\sqrt{T}} \leq \frac{\P(\tau_n \leq T)}{\sqrt{T}} = O(T^{pH - \frac{1}{2}})
    \]
    for an arbitrary large $p > 0$. Hence, using this estimate together with the one from Proposition \ref{prop: diff iv's} for the first term we have that, for every fixed $m \in \N$,
    \[
    \lim_{T \to 0} \frac{|\partial_kI (T,k^*) - \partial_kI^n(T,k^*)|}{T^m} = 0,
    \]
    as desired.
\end{proof}
We want to give a similar result for the curvature. In this case, since we want to avoid working with the density function of $X_T$, we will prove the following result using the call price representation in \cite{Romano97}.
\begin{prop} \label{prop: diff curvatures}
    Let $\mathcal{C}$ denote the price of the at-the-money call option under $V$ and let $\mathcal{C}^n$ denote the price of the at-the-money call option under $V^{n,n}$. Let $I^n$ and $I$ denote the implied volatilities for the models with the approximated and the original variance, respectively. Then, for $m \in \N$ large enough, we have
    \[
    \lim_{T \to 0} \frac{|\partial^2_{kk}I(T,k^*) - \partial^2_{kk}I^n(T,k^*)|}{T^m} = 0.
    \]
\end{prop}
\begin{proof}
    By differentiating the call price $\mathcal{C}$ with respect to $k$ twice, we obtain the following identity
    \begin{align*}
        \partial_{kk}^2\mathcal{C}|_{k = k^*} = &\partial^2_{kk}\BS(T,0,X_0,I(T,k^*)) \\
        + & 2\partial^2_{k\sigma} \BS(T,0,X_0,I(T,k^*))\partial_kI(T,k^*) \\
        + & \partial^2_{\sigma\sigma} \BS(T,0,X_0, I(T,k^*))(\partial_kI(T,k^*))^2 \\
        + & \partial_{\sigma}\BS(T,0,X_0,I(T,k^*)) \partial_{kk}^2I(T,k^*).
    \end{align*}
    Hence, by the asymptotic properties of the Black-Scholes function as $T \to 0$, Proposition \ref{prop: diff iv's} and Proposition \ref{prop: diff skews}, it is enough to prove
    \[
    \lim_{T \to 0} \frac{\partial_{kk}^2\mathcal{C}- \partial^2_{kk}\mathcal{C}^n}{T^m} = 0
    \]
    for $m \in \N$ arbitrarily large in order to conclude the result. Towards this end, we write the call prices $C$ and $C^n$ using a suitable representation. Conditioning to $W$ and $L$, the log-price is Gaussian. Hence, the call price can be expressed as
    \[
    \mathcal{C} = \E\left[ \BS(T,0, Z_T, \sqrt{1-\rho^2}v_0\right]
    \]
    where
    \[
    Z_T= X_0 + \rho\int_0^T \sqrt{V_s}dW_s - \frac{\rho^2}{2}\int_0^T V_s ds + L_T - \eta T
    \]
    and
    \[
    v_0^2 = \frac{1}{T}\int_0^T  V_s ds.
    \]
    We define $Z_T^n$ and $v_0^n$ in the same way as $Z_T$ and $v_0$ with the approximated variance process $V^{n,n}$. We also define $d_{-}$ and $d_{-}^n$ as
    \[
    d_{-}(k) = \frac{Z_T - k}{\sqrt{1-\rho^2}v_0\sqrt{T}} -\frac{\sqrt{1-\rho^2}v_0 \sqrt{T}}{2}, \quad d_{-}^n(k) = \frac{Z_T^n - k}{\sqrt{1-\rho^2}v_0^n\sqrt{T}} -\frac{\sqrt{1-\rho^2}v_0^n \sqrt{T}}{2}.
    \]
    Then, we have
    \[
    \partial_{kk}\mathcal{C} = e^{X_0}\E\left[\frac{\phi(d_{-}(k^*))}{v_0\sqrt{T}} -\Phi\left( d_{-}(k^*) \right) \right].
    \]
    Let $\tau_n = \inf\{t > 0 ; V_t \neq V_t^n\}$. Then
    \[
    |\partial_{kk}^2\mathcal{C}- \partial_{kk}^2\mathcal{C}^n| \leq e^{X_0}\left[\1_{\{\tau_n\leq T\}} \left( 2 + \frac{1}{\sqrt{2\pi T}v_0} + \frac{1}{\sqrt{2\pi T}v_0^n}\right) \right].
    \]
 In the event $\{\tau_n \leq T\}$ we have
    \[
    v_0^2T \geq (1-\rho^2) \frac{\tau_n}{n}, \quad v_0^n \geq (1-\rho^2) \frac{T}{n^2},
    \]
    because $V^{n,n} = \sigma_n(V^n)$ and $\sigma(x) \geq n^{-2}$ uniformly on $x \in \R$. Observe that for all $u \in [0,T]$
    \[
    \{\tau_n \leq u\} \subseteq \left\{\sup_{s \in [0,u]} |V_s^n- V_0| \geq \delta_n\right\},
    \]
    so $F_{\tau}(u) := \P(\tau_n \leq u) \leq C_{p,n} \delta_{n}^{-p} u^{pH}$ for all $p \geq 2$. Hence,
    \[
    |\partial^{2}_{kk} \mathcal{C} - \partial^2_{kk}\mathcal{C}^n| \leq C_{n,\rho}\left( \P(\tau_n \leq T) + \E[\tau_n^{-1/2} \1_{\{\tau_n \leq T\}}] + \frac{1}{\sqrt{T}}\P(\tau_n \leq T)\right).
    \]
    Since $\P(\tau_n \leq u) \leq C_{p,n} \delta_{n}^{-p} u^{pH}$ for an arbitrarily large $p \geq 2$, we only need to take care of the term $\E[\tau_n^{-1/2} \1_{\{\tau_n \leq T\}}]$. In this case, since $F_{\tau}$ is the cdf of $\tau_n$, we can integrate by parts to get
    \begin{align*}
   \E[\tau_n^{-1/2} \1_{\{\tau_n \leq T\}}] = &\int_0^T u^{-\frac{1}{2}}dF_{\tau}(u) = T^{-\frac{1}{2}}F_{\tau}(T) + \frac{1}{2}\int_0^T u^{-\frac{3}{2}}F_{\tau} (u) du.
    \end{align*}
    Hence,
    \[
    \E[\tau_n^{-1/2} \1_{\{\tau_n \leq T\}}]  \leq C_{p,n} \delta_n^{-p} \left( 1+\frac{1}{2pH+1}\right)T^{pH-\frac{1}{2}}.
    \]
    Since $p$ can be chosen arbitrarily large, for any fixed $m \in \N$
    \[
     \lim_{T \to 0} \frac{\partial_{kk}^2\mathcal{C} - \partial^2_{kk}\mathcal{C}^n}{T^m} = 0
    \]
    and, therefore, for any fixed $m \in \N$ (not necessarily the same $m$ as before) we have
    \[
     \lim_{T \to 0} \frac{\partial_{kk}^2I(T,k^*) - \partial^2_{kk}I^n(T,k^*)}{T^m} = 0,
    \]
    as desired.
\end{proof}
These results allow us to transfer the asymptotic behaviour from the approximated rough Heston model to the original rough Heston model as long as the approximated Heston model satisfies certain regularity conditions on the Malliavin derivatives. The next sections aim to show the Malliavin regularity of the process $V^n$. If satisfied, the Malliavin regularity for $V^{n,n}$ is ensured since $\sigma_n(x)$ is a regular function.

\subsection{Analysis of the first Malliavin derivative}
Since the coefficients $\sigma_n$ and $b_n$ are infinitely differentiable and bounded functions with $\sigma_n(x) \geq n^{-2}$ for all $x \in [0,\infty)$, it holds that $V^n \in \D^{1,2}$ and the Malliavin derivative of $V^n$ solves the equation
\[
D_sV_t^n = \nu K(t-s)\sqrt{\sigma_n(V^n_s)} - \kappa\int_s^t K(t-u) b_n'(V_u^n)D_sV^n_u  du + \nu\int_s^t K(t-u) \frac{\sigma_n'(V^n_u)}{2\sqrt{\sigma_n(V_u^n)}} D_sV^n_u dW_u
\]
In order to streamline the notation, we will name $Y_t = D_sV_t^n$, $a_u =  \frac{\sigma_n'(V^n_u)}{2\sqrt{\sigma_n(V_u^n)}}$ and $b_u = b_n'(V_u^n)$, so that $Y_t = D_sV_t^n$ solves the equation
\[
Y_t = \nu K(t-s)\sqrt{\sigma_n(V_s^n)} - \kappa\int_0^t K(t-u)b_u Y_udu + \nu\int_0^t K(t-u) a_u Y_u dW_u.
\]
We will write $|a| = \sup_{\omega \in \Omega}\sup_{u \in [s,t]} |a_u|$ to denote the supremum of the process $a$ (same notation applies to $b$).
\begin{prop} \label{prop: bound first derivative}
    There exists a constant $C = C_{p,n} > 0$ such that
    \[
    \E[|D_sV^n_t|^p]^{1/p} \leq C K(t-s).
    \]
\end{prop}
\begin{proof}
    From now on, $C$ and $C_j$ for $j \in \N$ will denote positive real constants that do not depend on $T$ and may switch from line to line. We can write
    \[
    Y_t = \nu K(t-s)\sqrt{\sigma_n(V_s)} + I_t + Z_t,
    \]
    where
    \[
    I_t = - \kappa\int_0^t K(t-u)b_u Y_udu, \quad Z_t = \nu\int_0^t K(t-u) a_u Y_u dW_u.
    \]
    Let $m(t) =||Y_t||_{L^p(\Omega)}$. We want to find a bound for $m(t)$. Regarding the term $I_t$, we can use Jensen's inequality with respect to the measure $K(t-u)du$ in order to obtain that
    \[
    ||I_t||_{L^p(\Omega)} \leq \int_s^t K(t-u)|b| ||Y_u||_{L^p(\Omega)} du \leq C\int_s^t K(t-u)m(u) du.
    \]
    Regarding the $Z_t$ term, the Burkholder-Davis-Gundy inequality together with Jensen's inequality allows us to write
    \[
    ||Z_t||_{L^p(\Omega)} \leq C\left(\int_s^t K(t-u)^2 |a|^2 ||Y_u||^2_{L^p(\Omega)} du\right)^{1/2} \leq C \left(\int_s^t K(t-u)^2 m(u)^2 du\right)^{1/2}.
    \]
    Now, since $\sigma$ is a bounded function and the map $t \mapsto \int_s^t K(t-u) du$ is increasing in $t$ we find that
    \[
    m(t) \leq C_1K(t-s) + C_2 \int_s^t K(t-u)m(u) du + C_3\left(\int_s^t K(t-u)^2 m(u)^2 du\right)^{1/2}.
    \]
    We define now the function
    \[
    M(t) = \sup_{u \in (s,t]} \frac{m(u)}{K(u-s)}.
    \]
    Since $m(u) \leq M(t)K(u-s)$ for all $u \leq t$ it holds that
    \[
    m(t) \leq C_1 K(t-s) + C_2 M(t) \int_s^t K(t-u)K(u-s) du + C_3 M(t) \left( \int_s^t K(t-u)^2 K(u-s)^2 du\right).
    \]
    Dividing by $K(t-s)$ and taking supremums, we find that
    \[
    M(t) \leq C_1 + M(t)(C_2A(t) + C_3B(t))
    \]
    where
    \[
    A(t) = \sup_{\tau \in (s,t]} \frac{\int_s^{\tau} K(\tau-u)K(u-s)du}{K(\tau-s)}, \quad B(t) = \sup_{\tau \in (s,t]} \frac{\left(\int_s^{\tau}K(\tau-u)^2 K(u-s)^2 du \right)^{1/2}}{K(\tau-s)}.
    \]
    On the one hand, there exists $C_H > 0$ such that
    \[
    \int_s^{\tau} K(\tau-u)K(u-s) du \leq C_H (\tau-s)^{2H},
    \]
    so there exists $C_H > 0$ such that $A(t) \leq C_H (t-s)^{H+1/2}$. Similarly, we can show that
    \[
    \left(\int_s^{\tau} K(\tau-u)^2 K(u-s)^2 du \right)^{1/2} \leq C_H(\tau-s)^{2H-1/2},
    \]
    so there exists $C_H > 0$ such that $B(t) \leq C_H (t-s)^{H}$. Since $|t-s|\leq T$, we have that
    \[
    M(t) \leq C_1 + M(t) (C_2 T^{H+1/2} + C_3 T^{H}) = : C_1 + M(t)\alpha_T.
    \]
    This inequality can also be written as
    \[
    M(t) \leq \frac{C_1}{1-\alpha_T}.
    \]
    Since $\alpha_T \to 0$ as $T \to 0$, we can assume without loss of generality that $T$ is small enough so that $\alpha_T \leq 1/2$ and therefore
    \[
    M(t) \leq 2C_1.
    \]
    Finally, for every $s \in [0,T]$ and for every $t \in [s,T]$ we have that
    \[
    ||Y_t||_{L^p(\Omega)} = m(t) \leq M(T) K(t-s) \leq 2C_1 K(t-s)
    \]
    concluding the desired result.
\end{proof}
We can, in fact, say a bit more about the first Malliavin derivative.
\begin{prop} \label{prop: first derivative}
    We have
    \[
    D_sV_t^n = \nu K(t-s)\sqrt{\sigma_n(V^n_s)} + R_{s,t}
    \]
    where $R_{s,t} =_{L^p(\Omega)} O(T^{2H-1/2})$.
\end{prop}
\begin{proof}
    We have
    \[
    R_{s,t} = -\kappa \int_s^t K(t-u)b_u D_sV^n_u du + \nu \int_s^t K(t-u)a_u D_sV^n_u dW_u =:A_1 + A_2.
    \]
    Since
    \[
    ||A_1||_{L^p(\Omega)} \leq C_n \int_s^t  K(t-u)||D_sV^n_u||_{L^p(\Omega)}du \leq C_n \int_s^t K(t-u)K(u-s) du 
    \]
    we have that
    \[
    ||A_1||_{L^p(\Omega)} \leq C_n T^{2H}.
    \]
    A similar argument for $A_2$ using Burkholder-Davis-Gundy inequality in \cite{Pannier2025} shows that $||A_2||_{L^p(\Omega)} \leq C_n T^{2H-1/2}$, concluding the result.
\end{proof}
\subsection{Analysis of the second Malliavin derivative}
The analysis of the second Malliavin derivative of $V^n$ is equivalent to the analysis of the first Malliavin derivative of $Y$. Again, since $\sigma_n$ and $b_n$ are smooth functions with $\sigma_n(x) \geq n^{-2}$ we have that $V^n \in \D^{2,2}$ and, moreover,
\begin{align*}
D_rD_sV_t^n = &\nu K(t-s) \frac{\sigma_n'(V_s)}{2\sqrt{\sigma_n(V_s)}}D_rV_s^n \\
- & \kappa\int_s^t K(t-u)b_n''(V_u^n) D_rV_u^n D_sV_u^n du \\
- & \kappa \int_s^t K(t-u) b'_n(V_u^n) D_rD_sV_u^n du \\
+ & \nu \int_s^tK(t-u) \left[ \frac{\sigma''_n(V_u^n)}{2\sqrt{\sigma_n(V_u)}} - \frac{\sigma'_n(V_u)^2}{4(\sigma_n(V_u^n))^{3/2}}\right]D_rV^n_u D_sV^n_u dW_u \\
+ & \nu \int_s^t K(t-u)\frac{\sigma'_n(V_u^n)}{2\sqrt{\sigma_n(V_u^n)}}D_rD_sV_u^n dW_u,
\end{align*}
provided $0 \leq r < s < t$. Again, we would like to streamline the notation. To this end, let $Y_{s,t} = D_sV_t^n$ and let $Z_{r,s,t} = D_rD_sV_t^n$. We also denote by $c, d$ the processes
\[
c_u = b''_n(V_u), \quad d_u = \frac{\sigma''_n(V_u^n)}{2\sqrt{\sigma_n(V_u)}} - \frac{\sigma'_n(V_u)^2}{4(\sigma_n(V_u^n))^{3/2}},
\]
so that we can write
\begin{align*}
    Z_{r,s,t} = & \nu K(t-s) a_s Y_{r,s} \\
    - & \kappa \int_s^t K(t-u) c_u Y_{r,u} Y_{s,u} du \\
    - & \kappa \int_s^t K(t-u) b_u Z_{r,s,u} du \\
    + & \nu \int_s^t K(t-u) d_u Y_{r,u} Y_{s,u} dW_u \\
    + & \nu \int_s^t K(t-u) a_u Z_{r,s,u} dW_u.
\end{align*}
\begin{prop} \label{prop: bound second derivative}
    There exists a constant $C > 0$ such that
    \[
    \E[|D_rD_sV_t|^p]^{1/p} \leq C K(t-s)K(s-r).
    \]
\end{prop}
\begin{proof}
Using the fact that $||Y_{s,t}||_{L^p(\Omega)} \leq C_p K(t-s)$ for all $p \geq 2$, we can apply Burkholder-Davis-Gundy inequality, Jensen's inequality, Hölder's inequality and the fact that $a,b,c,d$ are bounded processes to show that there exist $C_j > 0$, $j = 1,2,3,4,5$ such that
\begin{align*}
    ||Z_{r,s,t}||_{L^p(\Omega)} \leq &C_1 K(t-s) K(s-r) \\
    + &C_2 \int_s^t K(t-u) K(u-r) K(u-s) du \\
    + &C_3 \int_s^t K(t-u) ||Z_{r,s,u||_{L^p(\Omega)}} du \\
    + &C_4 \left(\int_s^t K(t-u)^2K(u-r)^2K(u-s)^2du\right)^{1/2} \\
    + &C_5 \left(\int_s^t K(t-u)^2 ||Z_{r,s,u}||^2_{L^p(\Omega)} du \right)^{1/2}
\end{align*}
Now, the same procedure as for the first Malliavin derivative can be done. We define the functions
\[
m(t) := ||Z_{r,s,u}||_{L^p(\Omega)}, \quad M(t) := \sup_{u \in (s,t]} \frac{m(u)}{K(u-s)K(s-r)}.
\]
Then, since $m(u) \leq M(t)K(u-s)K(s-r)$, we have
\begin{align*}
    M(t) \leq &C_1 +C_2 \frac{\int_s^t K(t-u)K(u-s)K(u-r) du}{K(t-s)K(s-r)} \\
    + &C_3 M(t)\frac{\int_s^t K(t-u)K(u-s)K(s-r) du}{K(t-s)K(s-r)} \\
    + & C_4 \frac{\left(\int_s^t K(t-u)^2 K(u-r)^2 K(u-s)^2 du \right)^{1/2}}{K(t-s)K(s-r)} \\
    + & C_5 M(t) \frac{\left( \int_s^t K(t-u)^2 K(u-s)^2 K(s-r)^2 du\right)^{1/2}}{K(t-s)K(s-r)}. \\
\end{align*}
Regarding the term multiplied by $C_2$, notice that the kernel $K$ is decreasing. Therefore, since $u-r = (u-s) + (s-r) \geq s-r$ we have that $K(u-r)\leq K(s-r)$. Hence,
\[
C_2\frac{\int_s^t K(t-u)K(u-s)K(u-r) du}{K(t-s)K(s-r)} \leq C_2\frac{\int_s^t K(t-u)K(u-s)K(s-r) du}{K(t-s)K(s-r)}  \leq C_2 T^{H+1/2}.
\]
This term has been easily estimated due to the cancellation of the $K(s-r)$ factor and the fact the resulting term is similar to the one that appeared in the analysis of the first Malliavin derivative. Notice that again, the term multiplied by $C_3$ can be simplified to get
\[
C_3M(t) \frac{\int_s^t K(t-u)K(u-s)K(s-r) du}{K(t-s)K(s-r)} = C_3 M(t) \frac{\int_s^t K(t-u)K(u-s) du}{K(t-s)}  \leq C_2 M(t)T^{H+1/2}.
\]
A similar bounding and cancelling argument show that
\[
M(t) \leq C_1+ C_2T^{H+\frac{1}{2}} +C_4 T^H + M(t)\left(C_3 T^{H + \frac{1}{2}} + C_5 T^H\right).
\]
Since $T \leq 1$ we can rename the constants to obtain a bound of the form
\[
M(t) \leq C_1 + M(t) \beta_T , \quad \beta_T = CT^H (\sqrt{T} + 1).
\]
Using that $\beta_T \to 0$ as $T \to 0$ we can conclude the result in the exact same way as with  the first Malliavin derivative. 
\end{proof}
The next result consists in finding the leading term of the second Malliavin derivative once we remove the singularity of the kernel.
\begin{prop}
    We have
    \[
    D_rD_sV_t^{n} = \nu K(t-s) \frac{\sigma_n'(V_s)}{2\sqrt{\sigma_n(V_s)}}D_rV_s^n + R_{r,s,t},
    \]
    where $R_{r,s,t} = O(T^{3H-1})$.
\end{prop}
\begin{proof}
    The proof follows the same strategy as in the proof of Proposition \ref{prop: first derivative}, using the bounds
    \[
    ||Z_{r,s,t}||_{L^p(\Omega)} \leq C (t-s)^{H-\frac{1}{2}}(s-r)^{H-\frac{1}{2}}, \quad ||Y_{s,t}||_{L^p(\Omega)} \leq C(t-s)^{H-\frac{1}{2}},
    \]
    together with Jensen's inequality and the Burkholder-Davis-Gundy inequality in \cite{Pannier2025} we obtain the result. We skip the details as the result is obtained via straightforward estimations and the procedure is exactly the same as in Proposition \ref{prop: first derivative}.
\end{proof}
\subsection{Analysis of the third Malliavin derivative}
Again, the analysis of the third Malliavin derivative of $V^n$ is completely analogous to the analysis of the first and second Malliavin derivatives. Let
\[
 Z_{r,s,t} = D_rD_sV_t^n, \quad Y_{s,t} = D_sV_t^n.
\]
Define the processes
\begin{align*}
e_u = b_n^{(3)}(V_u^n), \quad f_u := \frac{\sigma_n^{(3)}(V_u^n)}{2\sqrt{\sigma_n(V^n_u)}} -\frac{3\sigma'_n(V_u^n)\sigma''_n(V_u^n)}{4\sigma_n(V_u^n)^{3/2}} + \frac{3\sigma_n'(V_u^n)^3}{8\sigma_n(V_u^n)^{5/2}}
\end{align*}
Then, the third Malliavin derivative satisfies the following equation.
\begin{align*}
   D_qD_rD_sV_t^n = & \nu K(t-s)(d_sY_{q,s}Y_{r,s} + a_s Z_{q,r,s}) \\
    - & \kappa \int_s^tK(t-u)e_uY_{q,u}Y_{r,u}Y_{s,u}   du \\
    - & \kappa \int_s^t K(t-u)c_u Z_{q,r,u}Y_{s,u} du \\
    -&\kappa \int_s^t K(t-u)c_uZ_{q,s,u}Y_{r,u} du \\
    - & \kappa \int_s^t K(t-u)c_uZ_{r,s,u} Y_{q,u} du \\
    - &\kappa \int_s^t K(t-s)b_uU_{q,r,s,u} du \\
    + & \nu \int_s^t K(t-u)f_u Y_{q,u}Y_{r,u} Y_{s,u} dW_u \\
    + & \nu \int_s^t K(t-u)d_u Z_{q,r,u}Y_{s,u} dW_u \\
    + & \nu \int_s^t K(t-u) d_u Z_{r,s,u} Y_{q,u} dW_u \\
    + & \nu \int_s^t K(t-u) a_u U_{q,r,s,u} dW_u.
\end{align*}
We have the following result.
\begin{prop}
    There exists a constant $C > 0$ such that
    \[
    \E\left[| D_qD_rD_sV_t^n |^p \right]^{1/p} \leq CK(t-s)K(s-r)K(r-q).
    \]
\end{prop}
\begin{proof}
    The analysis of the third Malliavin derivative is obviously longer and more tedious than the analysis of the two other Malliavin derivatives. However, the computations are performed in the exact same way. We skipt it for the sake of conciseness.
\end{proof}
\subsection{The ATM-IV skew and curvature}
Here we gather all the results about $V^{n}:= V^{n,n}$ in order to find the explicit expression of the ATM-IV skew and curvature under the rough Heston model with stable-like jumps on the log-price. On the one hand, Hypothesis \ref{hyp 2} is satisfied by construction. Hypothesis \ref{hyp 4} is a consequence of Propositions \ref{prop: bound first derivative} and \ref{prop: bound second derivative} and, moreover, Hypothesis \ref{hyp 3} is implied by the fact that Hypothesis \ref{hyp 4} is satisfied. Hypothesis \ref{hyp 5} holds true by continuity of the sample paths and Hypothesis \ref{hyp 6} is derived from Proposition \ref{prop: first derivative}. Indeed, this constant $\Sigma$ can be found as
\[
\Sigma = \lim_{r,s \to 0} \frac{D_sV^n_r}{(r-s)^{H-1/2}}    = \frac{\nu \sigma_0}{\Gamma(H+\frac{1}{2})}.
\]
Hypotheses \ref{hyp 7} -- \ref{hyp 9} are also verified. Thus, we can compute the constant $\Theta$ as
\[
\Theta = \lim_{T \to 0} \frac{D_rD_sV_t^n}{(t-s)^{H-\frac{1}{2}}(s-r)^{H-\frac{1}{2}}} = \frac{\nu^2}{2\Gamma(H+\frac{1}{2})}.
\]Since the asymptotic behaviour of the ATM-IV skew and curvature under the rough Heston model with  stable-like jumps is the same as the asymptotic behaviour under the approximated version, we resort to Theorem \ref{th: main result} and Theorem \ref{th: main result 2} to determine that
    \[
    \lim_{T\to0}
    \frac{\partial_k I(T,k^*)}
    {T^{\min\left\{H-\frac12,\frac{1-Y}{2}\right\}}}
    =
    \begin{cases}
    \displaystyle
    \frac{\rho\nu}
    {2\sqrt{V_0}\left(H+\frac12\right)\left(H+\frac32\right)\Gamma\left( H+\frac{1}{2}\right)},
    & H<1-\frac{Y}{2},
    \\[2mm]
    \displaystyle
    \frac{(C_+-C_-)\Gamma\left(\frac{3-Y}{2}\right)}
    {V_0^{Y/2}Y(Y-1)2^{(Y-1)/2}},
    & H>1-\frac{Y}{2},
    \\[2mm]
    \displaystyle
    \frac{\rho\nu}
    {2\sqrt{V_0}\left(H+\frac12\right)\left(H+\frac32\right)\Gamma\left( H+\frac{1}{2}\right)}
    +
    \frac{(C_+-C_-)\Gamma\left(\frac{3-Y}{2}\right)}
    {V_0^{Y/2}Y(Y-1)2^{(Y-1)/2}},
    & H=1-\frac{Y}{2},
    \end{cases}
    \]
and
\[
\lim_{T \to 0} \frac{\partial^2_{kk}I(T,k^*)}{T^{\min\{2H-1, -\frac{Y}{2}\}}} = \begin{cases}
    \displaystyle
    C_{vol} & 2H-1 < -\frac{Y}{2}, \\
    \displaystyle
    C_{jumps} & 2H-1 > - \frac{Y}{2}, \\
    \displaystyle
    C_{vol} + C_{jumps} & 2H-1 = -\frac{Y}{2},
\end{cases}
\]
where 
\begin{align*}
C_{vol} = &\frac{\nu^2}{4V_0^{3/2}\Gamma(H+\frac{1}{2})^2(H+\frac{1}{2})^2} \\
& \times \left[ \frac{1}{2H+2} + \rho^2 \left(-\frac{6}{(H+\frac{3}{2})^2} + 4B\left( H + \frac{3}{2}, H + \frac{3}{2}\right) \right)\right],
    \end{align*}
    and
    \[
    C_{jumps} = \frac{(C_+ + C_-)(2-Y)\Gamma\left(1-\frac{Y}{2}\right)}{\sigma_0^{Y+1}Y (Y-1)2^{Y/2}}.
    \]
We choose the Lévy process $L$ such that it has Lévy measure
\[
\nu(dy) = |y|^{-1-Y} \left(C_+e^{-My}\1_{\{y>0\}} +  C_{-} e^{-G|y|}\1_{\{y < 0\}} \right).
\]
To compute the ATM-IV skew, let
\[
p_T=\P(X_T>X_0),
\qquad
q_T=\E\left[e^{X_T-X_0}\1_{\{X_T>X_0\}}\right].
\]
Following the representation in \cite{Heston1993}, the ATM call price $C$ can be written as
\[
C=S_0(q_T-p_T).
\]
Consequently, the ATM-IV $I(T,k^*)$ can be computed as
\begin{equation}\label{eq: ATM IV}
I(T,k^*)
=
\frac{2}{\sqrt{T}}
\Phi^{-1}\left(\frac{1+q_T-p_T}{2}\right).
\end{equation}
Moreover, the ATM-IV skew is given by
\begin{equation}\label{eq: ATM-IV skew}
\partial_k I(T,k^*)
=
\frac{
\Phi\left(-\frac{I(T,k^*)\sqrt{T}}{2}\right)-p_T
}{
\sqrt{T}\phi\left(\frac{I(T,k^*)\sqrt{T}}{2}\right)
}.
\end{equation}
Therefore, the computation of the ATM-IV skew reduces to that of $p_T$ and $q_T$. To compute these two quantities, we rely on the characteristic function of the log-return and Fourier inversion. Let $R_T=X_T-X_0$ and denote its characteristic function by
\[
\varphi_T(z)=\E\left[e^{izR_T}\right].
\]
We can write $R_T$ as
\[
R_T = X_T^{(c)} + L_T - \eta T.
\]
The inversion formula of \cite{gil1951note} yields
\[
p_T
=
\frac12+\frac{1}{\pi}
\int_0^\infty\frac{\operatorname{Im}(\varphi_T(u))}{u}\,du.
\]
Similarly, since
\[
e^{R_T}e^{iuR_T}=e^{i(u-i)R_T},
\]
we obtain
\[
q_T
=
\frac12+\frac{1}{\pi}
\int_0^\infty\frac{\operatorname{Im}(\varphi_T(u-i))}{u}\,du.
\]

It remains to determine $\varphi_T$. Since the rough Heston and Lévy components are independent, their characteristic functions factorize.  Let $\phy_T^{(c)}$ denote the characteristic function of $X_T^{(c)}$. Setting $\alpha=H+\frac12$, the affine transform formula of \cite{ElEuchRosenbaum2019} gives
\[
\varphi_T^{(c)}(z)
=
\exp\left(
V_0 I^{1-\alpha}h_z(T)
+\kappa\theta\int_0^T h_z(s)\,ds
\right),
\]
where $h_z$ solves the fractional Riccati equation
\[
h_z=I^\alpha F(z,h_z),
\qquad
F(z,x)
=
-\frac12(z^2+iz)
+(\rho\nu iz-\kappa)x
+\frac{\nu^2}{2}x^2,
\]
which is numerically solved using a fractional Euler scheme. For the CGMY component, \cite{CarrGemanMadanYor2002} obtain the characteristic exponent
\[
\chi(z)
=
\Gamma(-Y)\left[
C_+\big((M-iz)^Y-M^Y\big)
+
C_-\big((G+iz)^Y-G^Y\big)
\right].
\]
The characteristic function of $R_T$ is therefore
\[
\varphi_T(z)
=
\varphi_T^{(c)}(z)
\exp\left(
T\left[\chi(z)-iz\chi(-i)\right]
\right).
\]
In order to validate numerically the results obtained in the paper, we will conduct three experiments. We write the ATM-IV skew as
\[
\partial_kI(T,k^*) \approx AT^{\beta}.
\]
The first experiment consists in identifying the exponent $\beta$ in three scenarios:
\begin{enumerate}
    \item The rough volatility dominance scenario: we consider the rough Heston parameters
    \[
    (V_0, \nu, \kappa, \theta, \rho, H) =(0.04, 0.45, 1.5, 0.04, -0.7, 0.1)
    \]
    together with the general CGMY parameters
    \[
    (C_+, C_{-}, G, M, Y) = (0.015, 0.025, 5, 8, 1.5).
    \]
    \item The jump dominance scenario: we consider the rough Heston parameters
    \[
    (V_0, \nu, \kappa, \theta, \rho, H) =(0.04, 0.2, 1.5, 0.04, -0.1, 0.35)
    \]
    together with the general CGMY parameters
    \[
    (C_+, C_{-}, G, M, Y) = (0.015, 0.025, 5, 8, 1.5).
    \]
    \item The critical exponent case:  we consider the rough Heston parameters
    \[
    (V_0, \nu, \kappa, \theta, \rho, H) =(0.04, 0.45, 1.5, 0.04, -0.7, 0.25)
    \]
    together with the general CGMY parameters
    \[
    (C_+, C_{-}, G, M, Y) = (0.015, 0.025, 5, 8, 1.5).
    \]
\end{enumerate}
Notice that, in all cases $Y = 1.5$ and $H$ has been selected to satisfy one of the three conditions
\[
H < 1-\frac{Y}{2}, \quad H  = 1-\frac{Y}{2}, \quad H > 1-\frac{Y}{2}.
\]
The second experiment will be to determine the coefficient of the leading term $A$ in case when the power $\beta$ is known. To validate (from a mathematical point of view) the accuracy of the estimation of the coefficient of the leading term $\hat{A}$, we will assume that the power $\beta$ is the theoretical power. 

\begin{figure}[H]
    \centering

    \begin{minipage}[t]{0.48\textwidth}
        \centering
        \includegraphics[width=\linewidth]{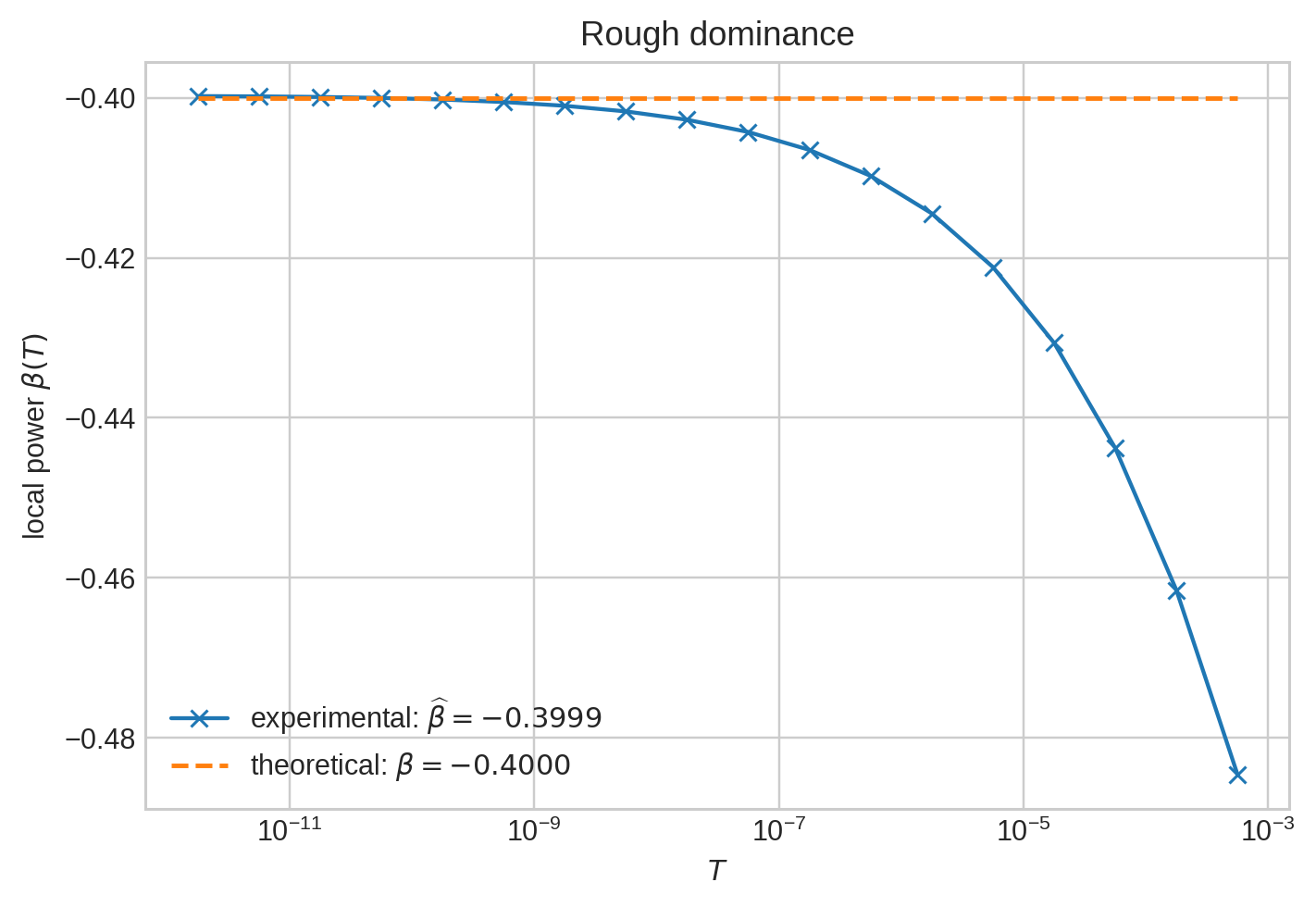}
    \end{minipage}
    \hfill
    \begin{minipage}[t]{0.48\textwidth}
        \centering
        \includegraphics[width=\linewidth]{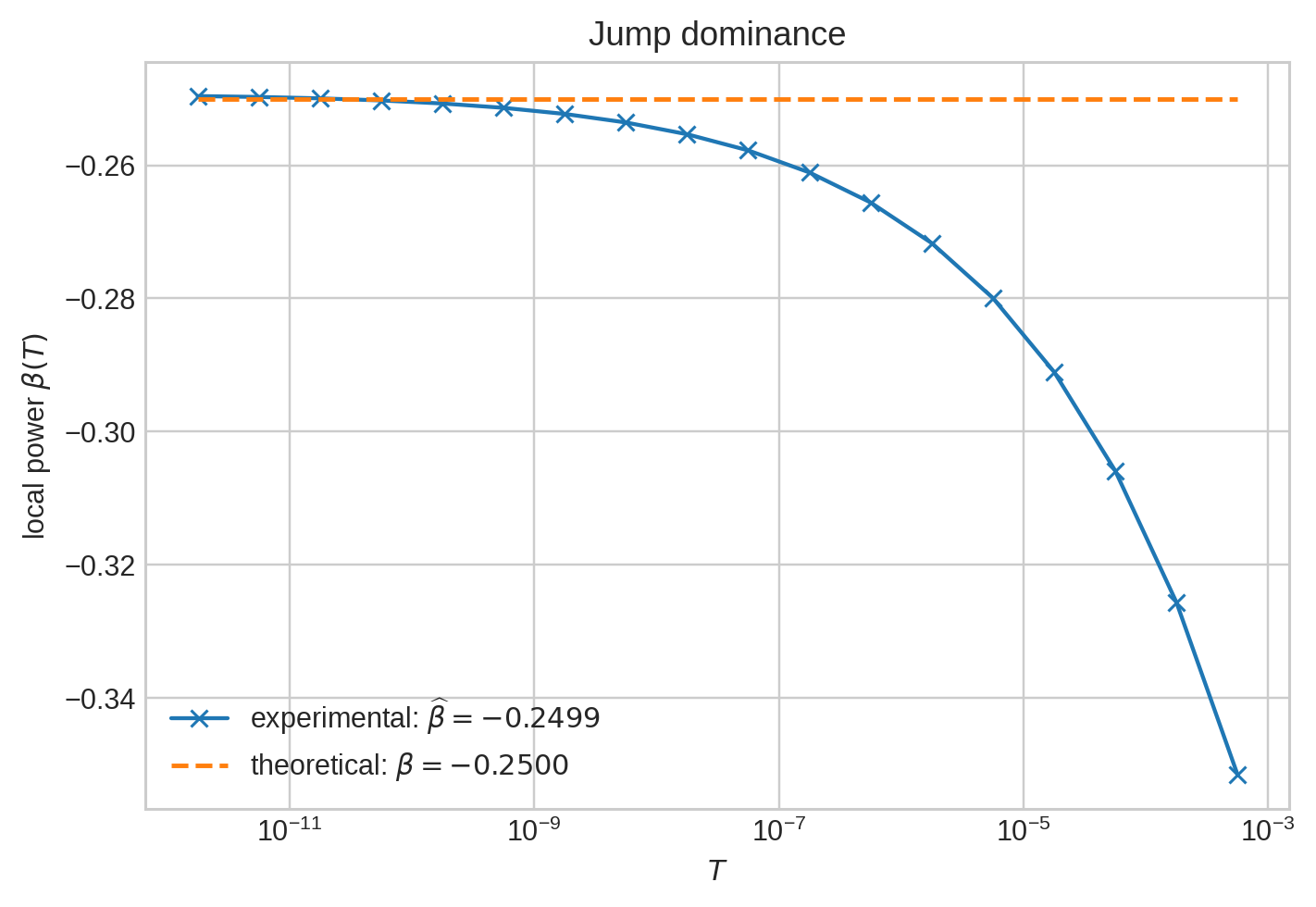}
    \end{minipage}

    \par\medskip

    \begin{minipage}[t]{0.48\textwidth}
        \centering
        \includegraphics[width=\linewidth]{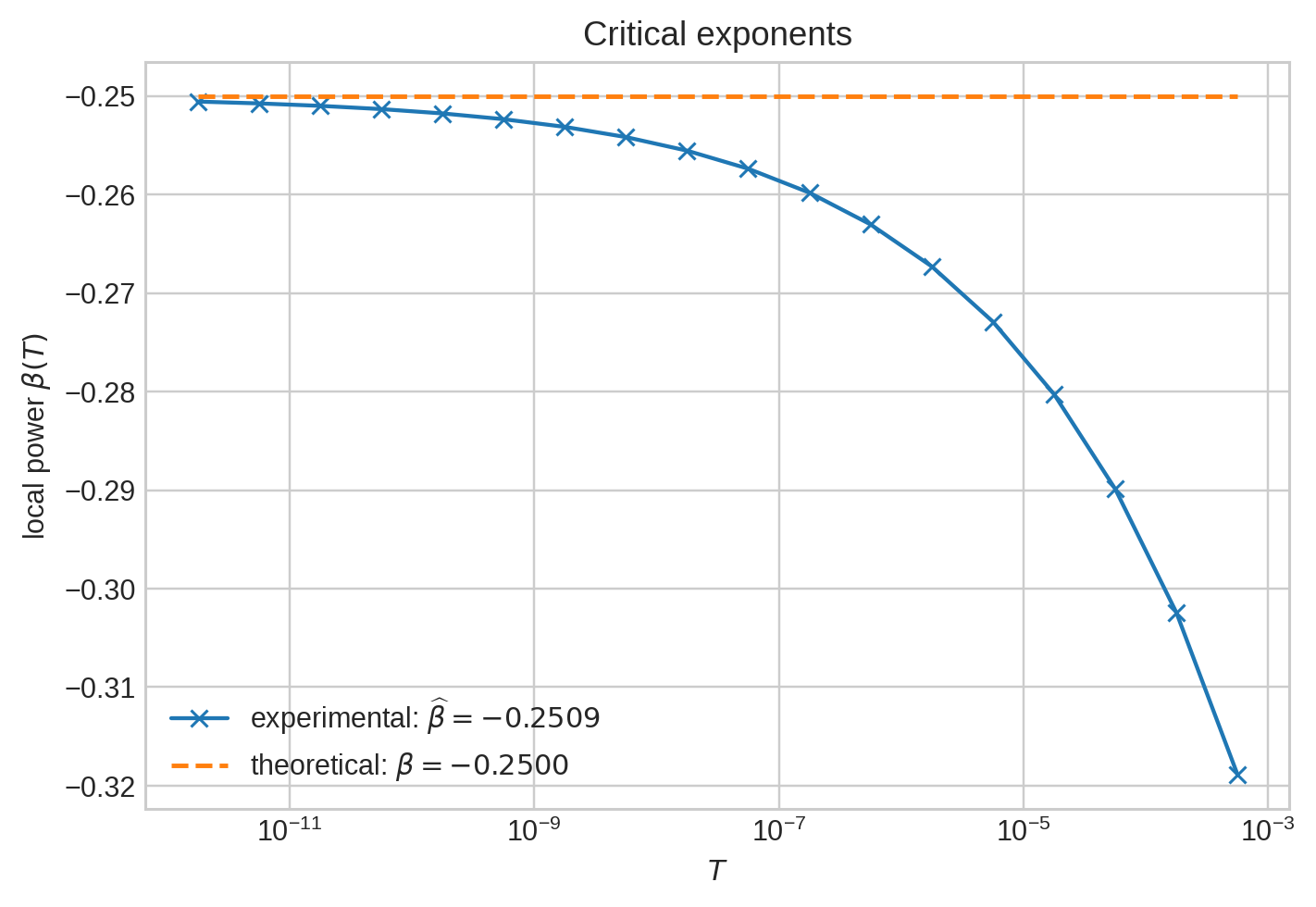}
    \end{minipage}

    \caption{Experimental vs. theoretical ATM-IV skew powers.}
    \label{fig:power-comparison}
\end{figure}

\begin{figure}[H]
    \centering

    \begin{minipage}[t]{0.48\textwidth}
        \centering
        \includegraphics[width=\linewidth]{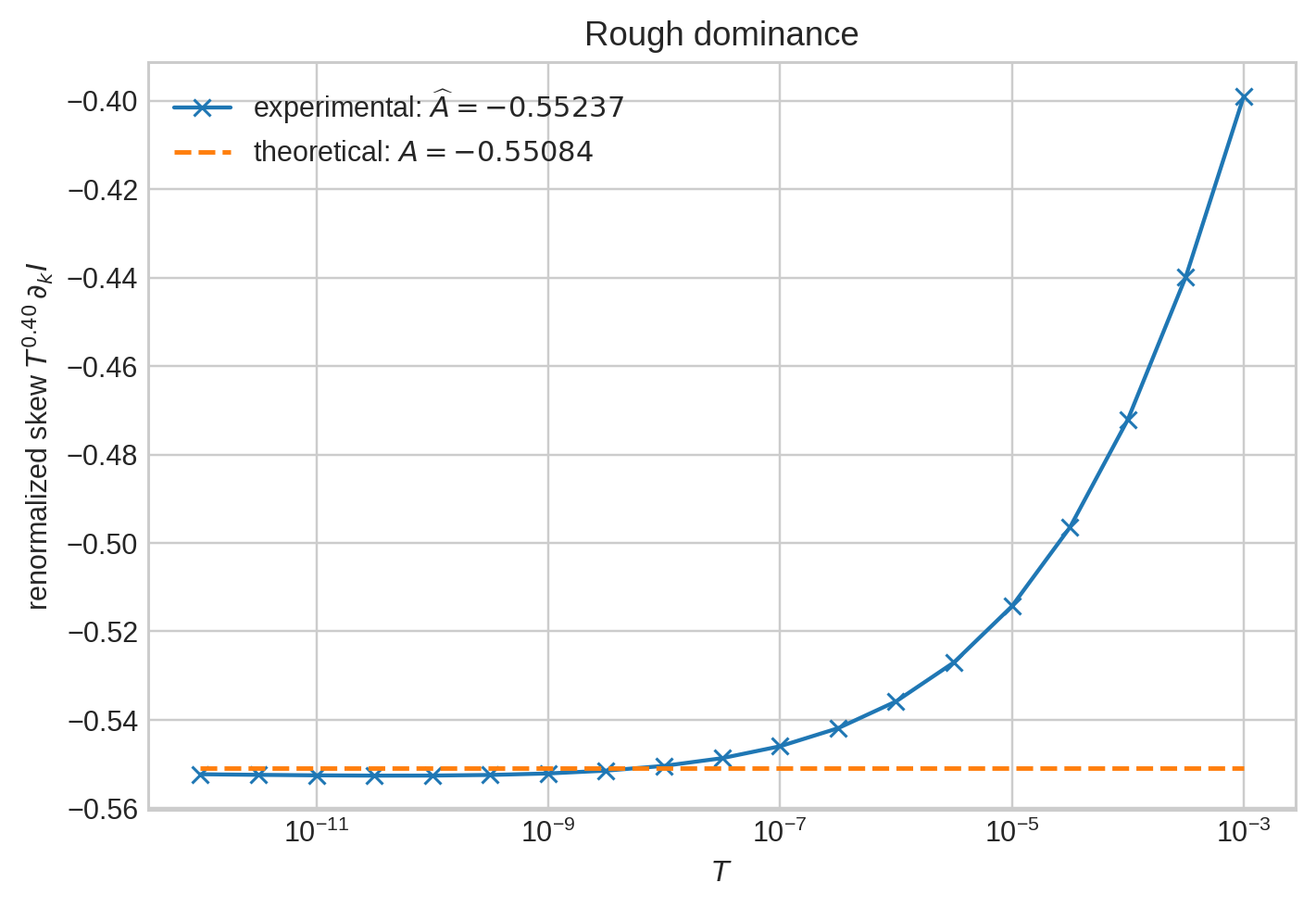}
    \end{minipage}
    \hfill
    \begin{minipage}[t]{0.48\textwidth}
        \centering
        \includegraphics[width=\linewidth]{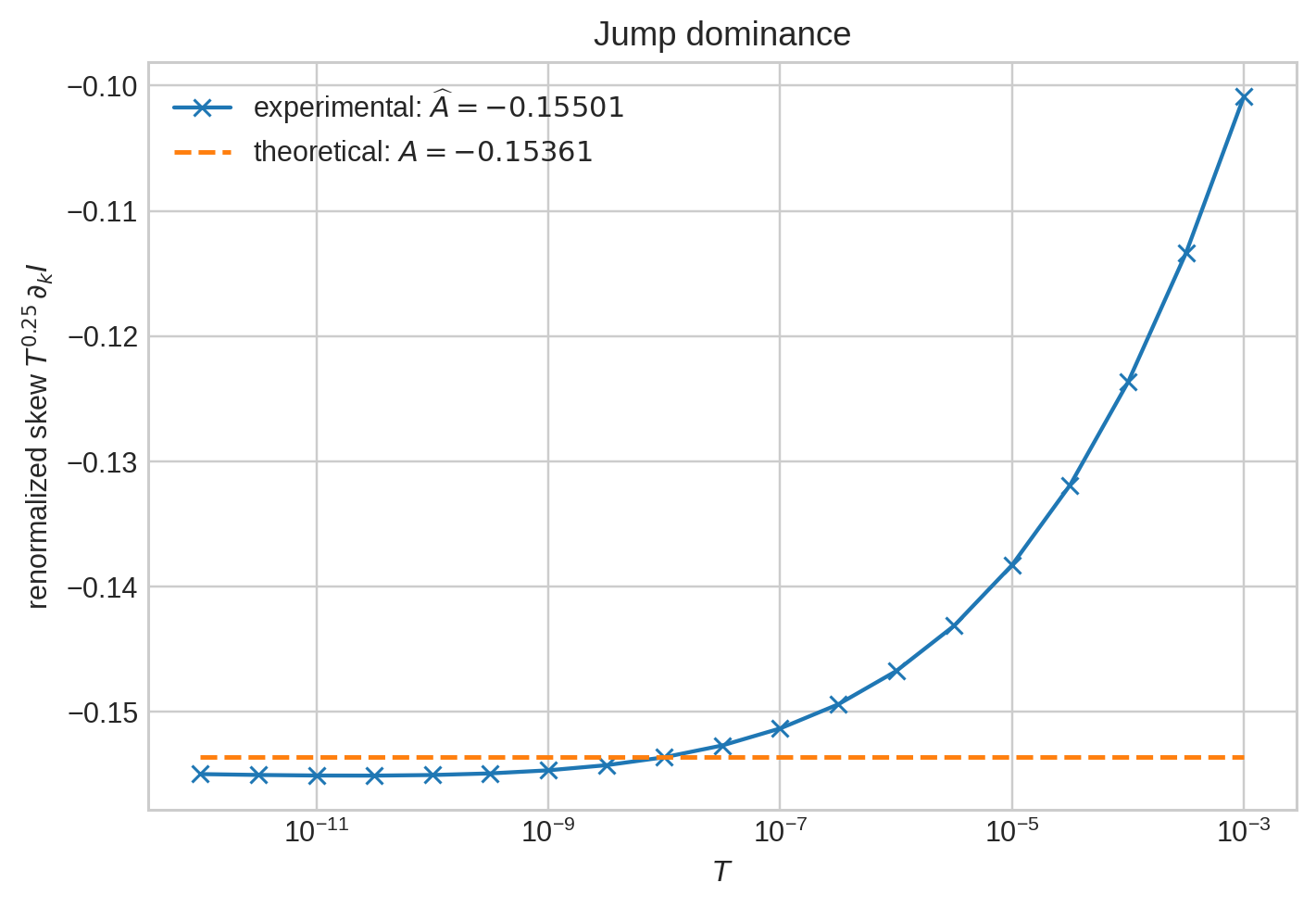}
    \end{minipage}

    \par\medskip

    \begin{minipage}[t]{0.48\textwidth}
        \centering
        \includegraphics[width=\linewidth]{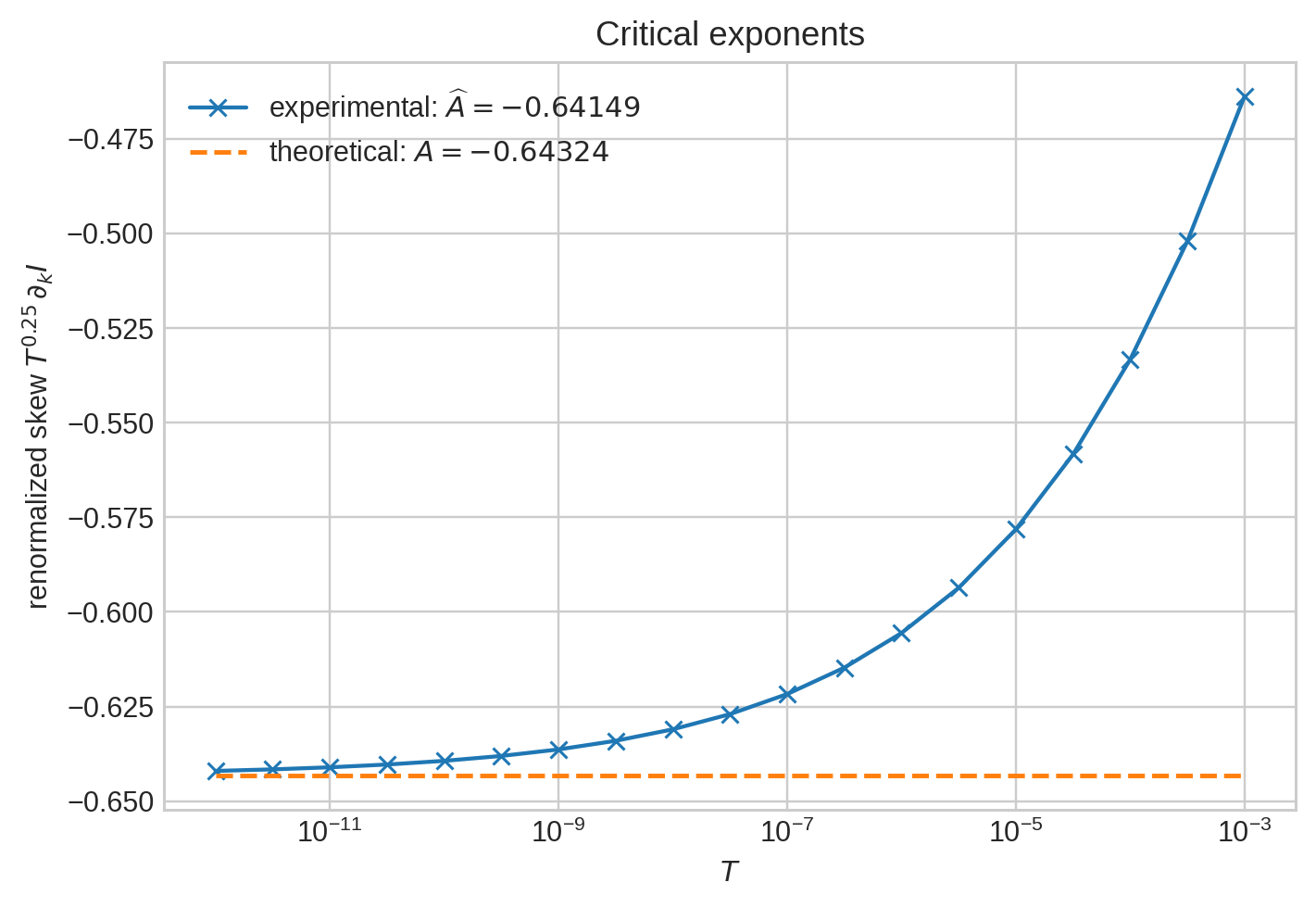}
    \end{minipage}

    \caption{Experimental vs. theoretical ATM-IV skew leading term coefficient.}
    \label{fig:coefficient-comparison}
\end{figure}

\begin{figure}[H]
    \centering

    \begin{minipage}[t]{0.48\textwidth}
        \centering
        \includegraphics[width=\linewidth]{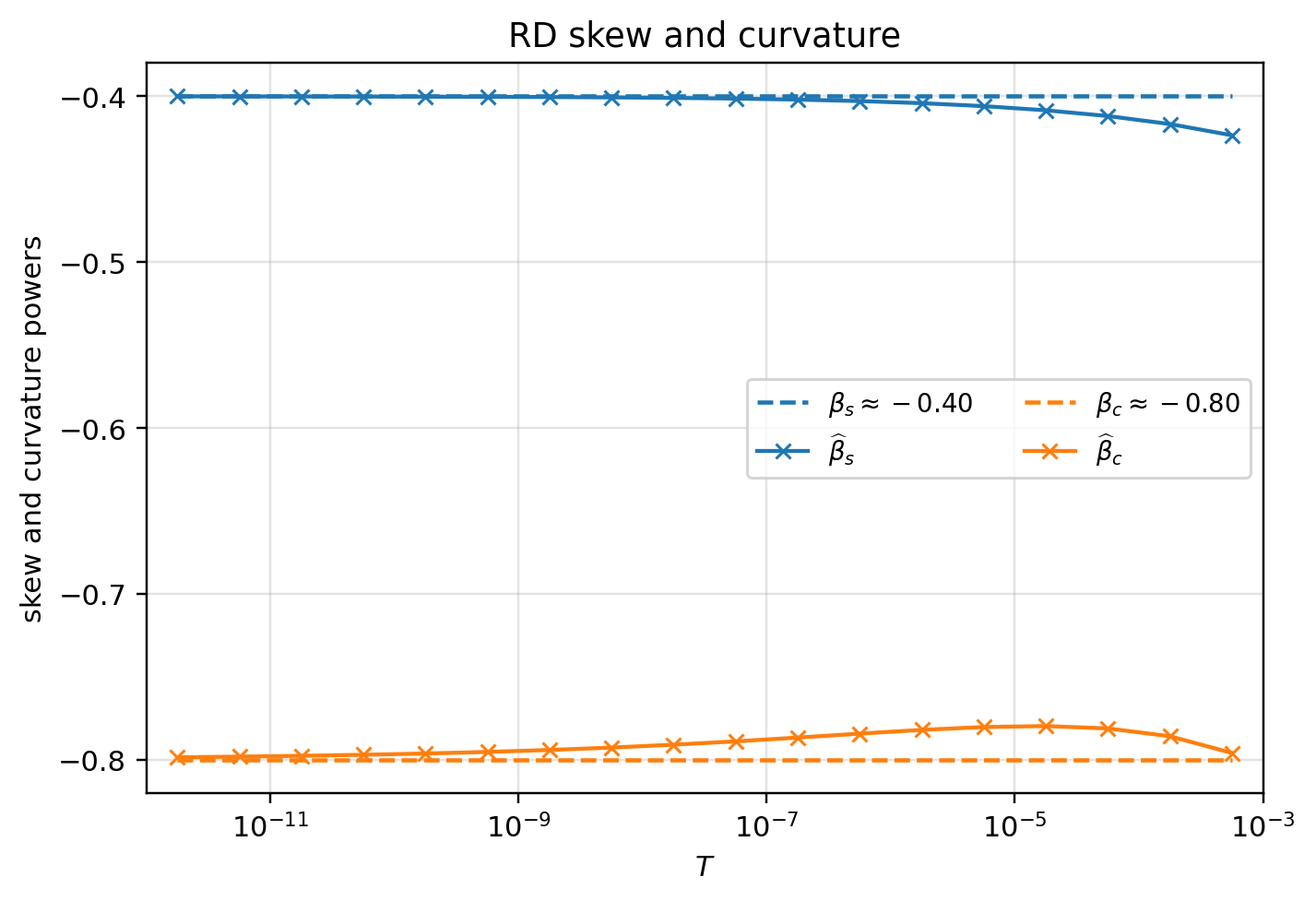}
    \end{minipage}
    \hfill
    \begin{minipage}[t]{0.48\textwidth}
        \centering
        \includegraphics[width=\linewidth]{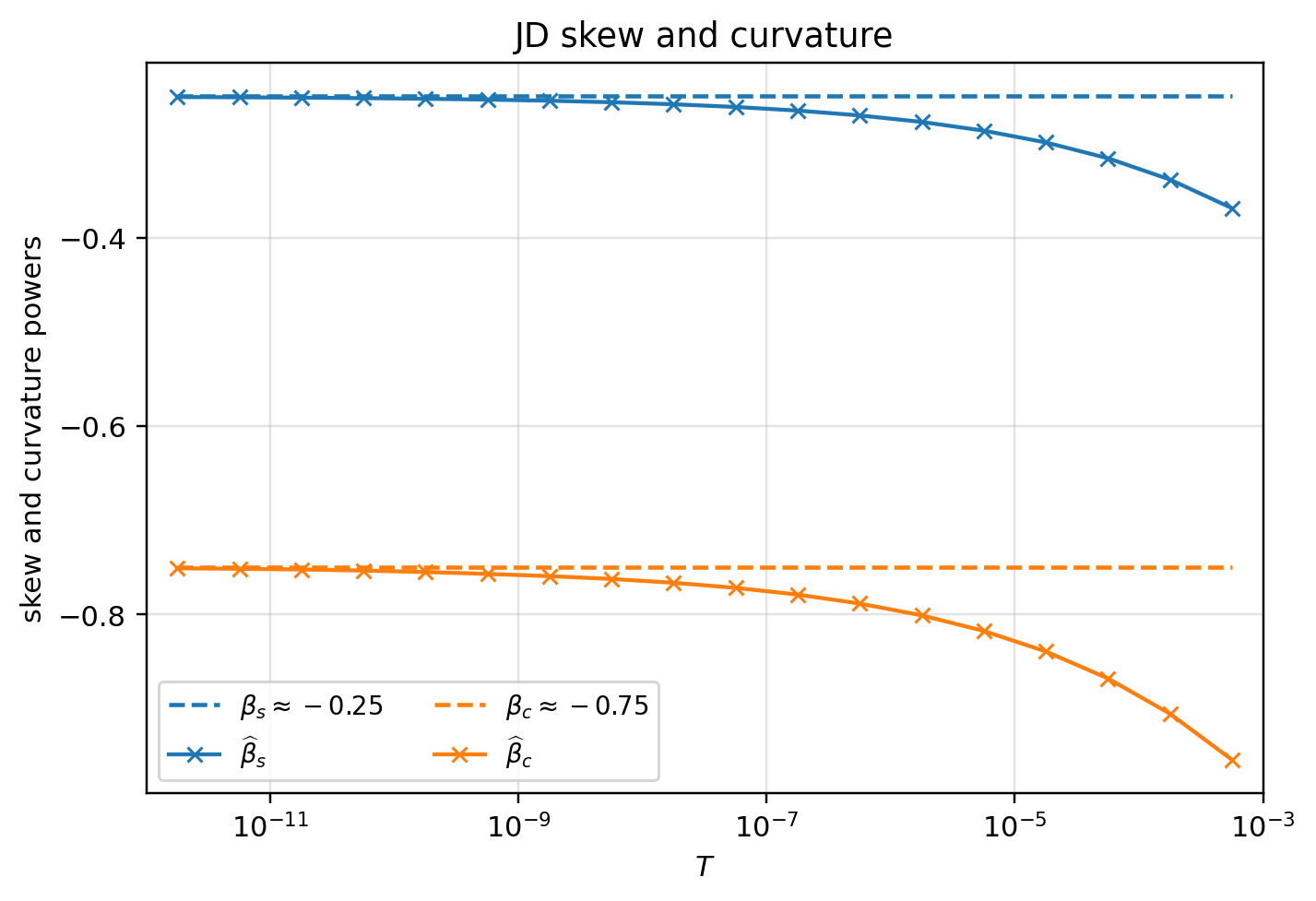}
    \end{minipage}

    \par\medskip

    \begin{minipage}[t]{0.48\textwidth}
        \centering
        \includegraphics[width=\linewidth]{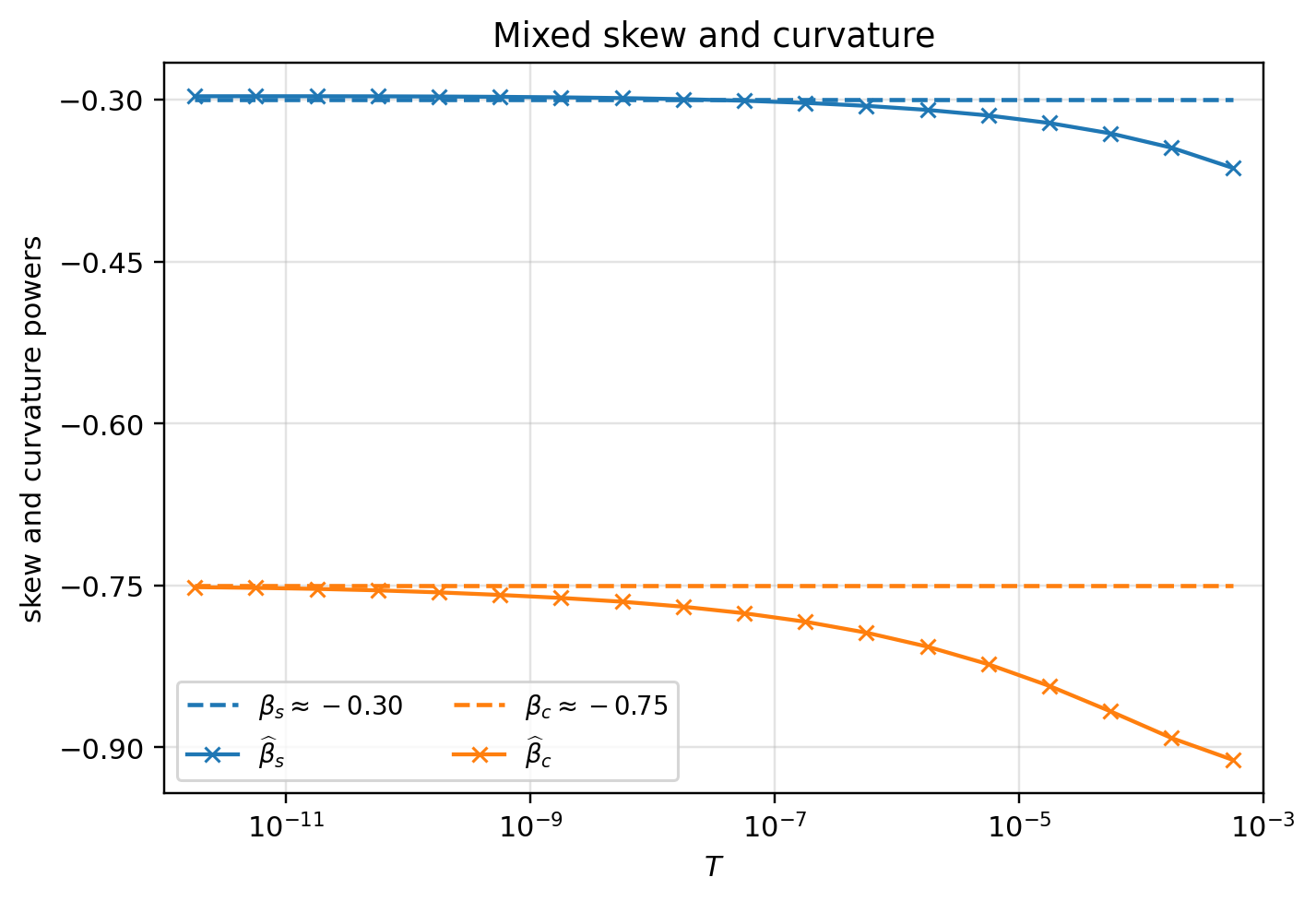}
    \end{minipage}

    \caption{Relationship between ATM-IV skew and curvature powers.}
    \label{fig: exponent comparison}
\end{figure}

In Figure \ref{fig:power-comparison} we see how the estimated power $\hat{\beta}$ approaches the theoretical skew power $\beta$ as $T \to 0$. In order to show that $\hat{\beta}$ is really close to $\beta$, we work with very small maturities that are not usually found in real markets. The reason why introduce such small maturities is for purely mathematical interest. Greater maturities, as seen in the plots, may generate ambiguity on the validity of our result.
Next, we want to see that the constant $A$ of the approximation $\partial_k I(T,k^*) \approx AT^{\beta}$ actually coincides in the three exponent regimes with the theoretical constant. In Figure \ref{fig:coefficient-comparison} we see again how the theoretical constant is attained for really small maturities, therefore validating the constants in Theorem \ref{th: main result}.
Regarding the curvature, we will show the the relationship between the ATM-IV skew and curvature exponents. Notice that the curvature can be computed as
\[
\partial^2_{kk}I(T,k^*) = \frac{\partial^2_{kk}\mathcal{C} - \partial^2_{kk}\BS -2\partial_{k\sigma}^2\BS \partial_kI(T,k^*) - \partial_{\sigma\sigma}^2\BS(T,\partial_kI(T,k^*))^2}{\partial_{\sigma}\BS},
\]
where we have used the abbreviation 
\[
\BS := \BS(T,0,X_0, I(T,k^*)).
\]
Since the method for computing $I(T,k^*)$ and $\partial_kI(T,k^*)$ has been described earlier, we only need to describe the computation of $\partial^2_{kk}C$. Denote by $f_{X_T}(x)$ the density function of $X_T$. Then
\[
\partial^2_{kk} \mathcal{C} |_{k = k^*} = e^{X_0}(f_{X_T}(0) -p_T).
\]
The density function is computed via Fourier inversion, that is,
\[
f_{X_T}(0) = \frac{1}{\pi} \int_0^{\infty} Re(\phy_T(u)) du.
\]
We will consider three scenarios:
\begin{enumerate}
    \item We consider $X$ following model \eqref{Black model} with rough Heston variance and without jumps, that is, $L \equiv 0$ and $\eta  = 0$. We consider the set of parameters
    \[
    (V_0, \kappa,\theta,\nu,H,\rho) = (0.04, 1.5, 0.04, 0.45, -0.5, 0.1).
    \]
    Since in this case the dynamics of the skew and curvature are purely driven by a rough volatility model, we will refer to this case as the rough dominated case, abbreviated as RD case.
    \item We consider $X$ following model \eqref{Black model} with constant volatility and with $L$ being a CGMY process. The parameters chosen are
    \[
    (\sigma, C_+, C_{-}, G, M, Y) = (0.2, 0.015, 0.025, 5, 8, 1.5).
    \]
    Now the dynamics are driven by a Lévy process. We will refer to this case as the jump dominated case, abbreviated as JD case.
    \item We consider $X$ following model \eqref{Black model} with a rough Heston variance and a CGMY process. The selected parameters are
    \[
       (V_0, \kappa,\theta,\nu,H,\rho) = (0.04, 1.5, 0.04, 0.45, -0.7, 0.2),
    \]
    and
    \[
    (C_+, C_{-}, G, M, Y) = (0.2, 0.015, 0.025, 5, 8, 1.5).
    \]
\end{enumerate}
We let $\partial_k I(T,k^*) \sim A_s T^{\beta_s}$ and $\partial_{kk}^2 I(T,k^*) \sim A_c T^{\beta_c}$. In the RD case, the theoretical powers of the skew and curvature are
\[
\beta_s = H- \frac{1}{2} = -0.4, \quad \beta_c = 2\beta_s = -0.8.
\]
In a similar manner, in the JD case, the theoretical powers of the skew and curvature are
\[
\beta_s = \frac{1-Y}{2} = -0.25, \quad \beta_c =\beta_s - \frac{1}{2} = -0.75.
\]
In the mixed scenario, since
\[
H-\frac{1}{2}<\frac{1-Y}{2}, \quad 2H-1 > -\frac{Y}{2}
\]
we break the relationship between skew and curvature powers. In this case, we have
\[
\beta_s = H-\frac{1}{2}= -0.3, \quad \beta_c = - \frac{Y}{2} = -0.75.
\]
In Figure \ref{fig: exponent comparison} we see how the experimental powers $\hat{\beta}_s$ and $\hat{\beta}_c$ approach the theoretical values as $T \to 0$ in the three regimes. 
The results of the approximation for the smaller maturity are displayed in Table \ref{tab:powers skew curv}.
\begin{table}[H]
    \centering
    \begin{tabular}{|c|c|c|c|c|}
    \hline
         & $\beta_s$  & $\hat{\beta}_s$  & $\beta_c$ & $\hat{\beta}_c$ \\
         \hline
        RD case & $-0.40$ & $-0.4001$ & $-0.80$ & $-0.7978$\\ 
        \hline
         JD case & $-0.25$  & $-0.2514$ & $-0.75$ & $-0.7520$\\
         \hline
         Mixed case& $-0.3$ & $-0.2969$ & $-0.75$ &$-0.7531$\\
         \hline
    \end{tabular}
    \caption{Dominating skew and curvature powers in the RD, JD and mixed cases.}
    \label{tab:powers skew curv}
\end{table}
\section{Conclusions}
In this paper, we have derived the leading short-time asymptotics of the ATM implied volatility skew and curvature in a model combining general stochastic volatility with a stable-like Lévy process. The result allows us to identify the mechanism generating the power law in the skew and curvature by analysing the relationship between the leading powers. To the best of our knowledge, this is the first result of this type covering both Markovian and rough volatility specifications together with such a general Lévy measure.

The proofs of the main results are obtained by a flexible methodology based on Malliavin calculus. Moreover, through an approximation argument, we extend the result to the rough Heston model despite the failure of some of the regularity assumptions of the general theorem. The numerical experiments support the theoretical asymptotics. We believe that this methodology can be further developed to obtain higher-order expansions and to study other short-time properties of the implied volatility surface.

\section*{Acknowledgements}
Author Òscar Burés is supported by program AGAUR-FI ajuts (2025 FI-1 00580) from the Department of Research and Universities of the Government of Catalonia and the co-funding of the European Social Fund Plus (ESF+).
\bibliographystyle{apalike}
\bibliography{references.bib}

\end{document}